\documentstyle[12pt]{article}

\advance\textheight by \topskip
\newtheorem{theorem}{Theorem}[section]

\newtheorem{lemma}[theorem]{Lemma}
\newtheorem{remark}[theorem]{Remark}
\newtheorem{coroll}[theorem]{Corollary}
\newtheorem{definition}[theorem]{Definition}
\newtheorem{example}[theorem]{Example}

\newcommand{\parfrac}[2]{\frac{\partial #1}{\partial #2}}
\newcommand{\p}{\partial}

\newcommand{\dps}{\displaystyle}
\newcommand{\eop}{\hspace*{\fill} $\Box$}
\newcommand{\proof}{\paragraph{{\it Proof.}}}
\newcommand{\wt}[1]{\widetilde{#1}}
\newcommand{\para}[1]{\paragraph{#1}}

\newcommand{\be}[1]{\begin{equation} \label{#1}}
\newcommand{\ee}{\end{equation}}

\def\e{\epsilon}
\def\d{\delta}
\def\g{{\cal G}}
\def\gedl{{\cal G}^{\epsilon  \delta n}}
\def\gdnt{{\cal G}^{\delta n}_{(0,T)}}
\def\ge{{\cal G}^{\epsilon }}
\def\tedl{\tau ^{\epsilon \delta n}}
\def\te{\tau ^{\epsilon }}
\def\R{{\rm I\! R}}
\def\N{{\rm I\! N}}
\def\charfct{1\! \mbox{l}}
\def\qm{quantum mechanics}
\def\conf{\R ^{d}}

\def\P{{\bf P}}
\def\vpsi{v^\psi }
\def\vpsit{v^{\psi _t}}
\newcommand{\wf}{wave function}
\newcommand{\BM}{Bohmian mechanics}
\newcommand{\sa}{self-adjoint}
\newcommand{\SA}{self-adjointness}
\newcommand{\ham}{Hamiltonian}
\newcommand{\wfs}{wave functions}
\newcommand{\roet}{\mbox{${\rho}^{\epsilon}_t$}}

\newcommand{\roedl}{\mbox{${\rho}^{\epsilon \delta n}$}}
\newcommand{\Schr}{Schr\"{o}dinger}
\newcommand{\seq}{Schr\"{o}dinger's equation}
\newcommand{\Qet}{\mbox{${Q}^{\epsilon}_t$}}
\newcommand{\Qedlt}{\mbox{${Q}^{\epsilon \delta n}_t$}}
\newcommand{\Qedl}{\mbox{${Q}^{\epsilon \delta n}$}}

\newcommand{\Qetq}{\mbox{${Q}^{\epsilon}_t(q)$}}
\def\No{{\cal N}}

\newenvironment{deflist}[1]%
{\begin{list}{}%
{\settowidth{\labelwidth}{#1}%
\setlength{\leftmargin}{\labelwidth}%
\addtolength{\leftmargin}{\labelsep}%
\setlength{\parsep}{0pt}%
\setlength{\itemsep}{0pt}%
\setlength{\topsep}{0pt}%
}}%
{\end{list}}%

\begin{document}
\noindent
{\Huge {\bf On the Global Existence}}

\vspace{4mm}

\noindent
{\Huge {\bf of Bohmian Mechanics}}

\vspace{15mm}

\noindent
K. Berndl$^1$,  D. D\"{u}rr$^1$, S. Goldstein$^2$, G. Peruzzi$^3$,
and
N. Zangh\`{\i}$^4$
\bigskip
{\footnotesize \begin{deflist}{8}
\item[1]
Mathematisches Institut der Universit\"{a}t M\"{u}nchen,\\
Theresienstra{\ss}e 39,
80333 M\"{u}nchen, Germany
\item[2] Department of Mathematics,
Rutgers University, New Brunswick, NJ 08903, USA
\item[3] Sezione INFN Firenze, Largo Fermi, 2, 50125 Firenze,
Italy
\item[4] Dipartimento di Fisica,
Universit\`a di Genova, Sezione INFN Genova,\\ Via Dodecaneso 33,
16146 Genova, Italy
\end{deflist}}
\medskip
\noindent
August 9, 1994

\bigskip

\bigskip

\para{Abstract.}  We show that the particle motion in Bohmian
mechanics, given by the solution of an ordinary differential equation,
exists globally: For a large class of potentials the singularities of the
velocity field and infinity will not be reached in finite time for typical
initial values. A substantial part of the analysis is based on the
probabilistic significance of the quantum flux. We elucidate the connection
between the conditions necessary for global existence and the
self-adjointness of the Schr\"odinger Hamiltonian.

\bigskip

\section{Introduction}

Bohmian mechanics \cite{Bohm52,BohmIII,Bell,DGZ1,DGZ3,Holland} is a Galilean
invariant theory for
the motion of point particles. Consider a system of $N$ particles with
masses $m_{1},...,m_{N}$ and potential $V=V({\bf Q}_1, \dots , {\bf Q}_N)$,
where ${\bf Q}_k \in \R ^{\nu}$ denotes the position of the $k$-th
particle. The relevant configuration space is an open subset of
${\nu}N=d$-dimensional space $\R ^d$, for example the complement of the set
of singularities of $V$, and shall be denoted by $\Omega$. The state of the
$N$-particle system is given by the configuration $ Q=({\bf Q}_1, \dots ,
{\bf Q}_N) \in \Omega$ and the Schr\"odinger \wf \ $\psi$ on configuration
space $\Omega$.  On the subset of $\Omega$ where the \wf\ $\psi \neq 0$ and
is differentiable, it generates a velocity field $v^{\psi}=({\bf
v}^{\psi}_{1},
\dots ,{\bf v}^{\psi}_{N})$
\be{velfield}
{\bf v}_{k}^{\psi}(q) = \frac{\hbar}{m_{k}}
\mbox {Im} \frac{{\nabla}_{k}\psi (q)}{\psi (q)}
\ee
the integral curves of which are the trajectories of the particles.
Thus the time evolution of the state $(Q_t,\psi _t)$ is given by a first-order
ordinary differential equation for the configuration $Q_t$
\begin{equation} \label{bohmev}
\frac {dQ_t}{dt} = \vpsit (Q_t)
\end{equation}
and \Schr 's equation for the \wf \ $\psi _t$
\begin{equation}  \label{seq}
i\hbar\frac{\partial \psi _t(q)}{\partial t}  = \left( -{\sum}_{k=1}^{N}
\frac{{\hbar}^{2}}{2m_{k}}\Delta_{k} + V(q)            \right)
 \psi _t(q),\end{equation}
where $\nabla _k$ and $\Delta _k$ denote the gradient and the Laplace
operator in $\R ^{\nu}$ and the potential $V$ is  a real-valued function on
$\Omega$.

\BM\ may be regarded as a fundamental nonrelativistic quantum theory, from
which the quantum formalism---operators as observables, the uncertainty
principle, etc.---emerges as
``measurement'' formalism.
It resolves all problems associated with the measurement problem
in \qm\ \cite{Bohm52,BohmIII,Bell,DGZ1,DGZ3,Holland}. It accounts for the
``collapse'' of the \wf , for
quantum randomness as expressed by Born's law $\rho = |\psi |^2$,
and familiar (macroscopic) reality. For a thorough analysis of the
physics entailed by \BM\ see \cite{Bohm52,BohmIII,DGZ1,DDGZ},
and \cite{DGZ3} for a short overview of \cite{DGZ1}.

Here we are concerned with the mathematical problem of the existence and
uniqueness of the motion in \BM, i.e., with establishing that for given
$Q_0$ and $\psi _0$ at some ``initial'' time $t_0$ ($t_0 = 0$), solutions
$(Q_t,\psi_t)$ of (\ref{bohmev}, \ref{seq}) with $Q_{t_0} = Q_0$ and $\psi
_{t_0} = \psi_0$ exist uniquely and globally in time. (Note that
Schr\"odinger's equation (\ref{seq}) is independent of the particle motion,
while for solving the Equation (\ref{bohmev}) for the particle motion we
need the \wf\ $\psi _t$.)  Our first motivation for addressing this problem
is the fact that the velocity field (\ref{velfield}) exhibits rather
obviously possible catastrophic events for the motion: $\vpsi$ is singular
at the nodes of $\psi$, i.e., at points where $\psi=0$, so that the
solution would break down if a node were reached. Furthermore, the solution
may cease to exist at singularities of the \wf\ (if it has singularities),
at the boundary of $\Omega$ (if it has a
boundary), and because of ``explosion,'' that is the escape to
infinity of a particle in a finite amount of time---events which have
analogues in the $N$-body problem (of gravitational interaction) in
Newtonian mechanics.

Recall that the problem of the existence of dynamics in Newtonian mechanics
is notoriously difficult \cite{Moser,Diacu}. In addition to the possibility
of routine collision singularities, the $N$-body problem with $N> 3$ yields
marvelous scenarios of so-called pseudocollisions, where some particles,
while oscillating wildly, reach infinity in finite time. Examples of such
catastrophies have been constructed by Mather and McGehee
\cite{MM},\footnote{However, this example,
which is 1-dimensional, involves an
infinite number of binary collisions before the system explodes and thus
does not describe a true pseudocollision.} by Gerver \cite{Gerver}, and by Xia
\cite{Xia}.  While, for the case of
a
``solar system'' with small ``planetary'' masses, Arnold \cite{Arnold}
established global existence (and much more) ``for the majority of initial
conditions for which the eccentricities and inclinations of the Kepler
ellipses are small,'' and while Saari \cite{Saari} has established global
existence for ``almost all initial conditions (in the sense of Lebesgue
measure and Baire category)'' for the 4-body problem, for systems of more
than four particles it is not known whether the initial conditions leading
to such catastrophies are atypical, i.e., form a set of Lebesgue measure
zero---though this is certainly expected by most experts to be the case
\cite{Diacu} (though not by all \cite{Mather}).
Indeed, apart from obvious scenarios---such as the particles moving apart
sufficiently rapidly---and those covered by some version of the KAM theorem
\cite{Arnoldbuch}, for $N\geq5$
it cannot, so far as we know, even be precluded on the
basis of what has so far been proven that this set has full measure!

It is remarkable that the situation in the corresponding quantum system is very
different. In orthodox quantum
theory the  time evolution of the state $\psi_t$ is given by a one-parameter
unitary group $U_{t}$ on a Hilbert
space $\cal H$. $U_t$ is generated by a self-adjoint operator $H$, which on
smooth \wfs\ in ${\cal
H}=L^{2}(\Omega )$ is given by
\begin{equation} \label{hamiltonian}
H = -{\sum}_{k=1}^{N}
\frac{{\hbar}^{2}}{2m_{k}}\Delta_{k} + V = H_0 + V,
\end{equation}
i.e., \Schr 's equation is regarded as the ``generator equation'' for
$U_t$. Hence the ``problem of the existence of dynamics'' for \seq\ is
reduced to that of showing that the relevant Hamiltonian $H$ (given by the
particular choice of the potential $V$) is self-adjoint. This has been done
in great generality, independent of the number of particles and for large
classes of potentials, including singular potentials like the Coulomb
potential, which is of primary physical interest \cite{kato51,ReeSiII}. We
shall discuss the meaning and the status of the self-adjointness of the
\ham\ from the perspective of \BM\ in Section 4. It may be worthwhile to
note, however, that the sufficiency of establishing only the \SA\ of the
\ham\ for a satisfactory physical interpretation has been questioned by
Radin and Simon \cite{RadSi}: ``Interestingly enough, while Kato's result
`solves' the dynamical existence question in the quantum case, it says
nothing about the question of ${\bf x}(t)^{2}$ remaining finite in time! {}From
its physical interpretation, proof of
such regularity property is
clearly desirable.''

In \BM\ we have not only \seq\ (\ref{seq}) to consider, but also the
differential equation (\ref{bohmev}),
governing the motion of the particles. Thus the question of existence of the
dynamics of \BM\ draws again nearer
to the situation in Newtonian mechanics,
as it depends now on detailed regularity properties of the velocity field
$\vpsi$ (\ref{velfield}). {\em Local}\/ existence and uniqueness of Bohmian
trajectories is guaranteed if the
velocity field $\vpsi$ is locally Lipschitz continuous. We therefore certainly
need greater regularity for the \wf\
$\psi$ than merely that $\psi$ be in $L^2(\Omega )$.

Global existence is more delicate. In addition to the nodes of $\psi$, there
are singularities  comparable to those of Newtonian mechanics. Firstly,
even for a globally smooth velocity field the solution of (\ref{bohmev})
may explode, i.e., it may reach infinity in finite time. Secondly, the
boundary points of $\Omega$, typically the singular points of the
potential, are reflected in singular behavior of the \wf\ at such points,
giving rise to singularities in the velocity field
(\ref{velfield}).\footnote{For example, the ground state \wf\ of one
particle in a Coulomb potential $V(q)=1/|q|$, $q\in \R ^3$ (``hydrogen
atom'') has the form $e^{-|q|}$, which is not differentiable at the point
$q=0$ of the potential singularity.}

The problem we address is the following: Suppose that at some arbitrary
``initial time'' ($t_0=0$) the $N$-particle configuration lies in the
complement of the set of nodes and singularities of $\psi _0$. Does the
trajectory develop in a finite amount of time into a singularity of the
velocity field $\vpsi$, or does it reach infinity in finite time?
According to Theorem \ref{mainth} and Corollary \ref{maincor}, the answer
is negative for ``typical'' initial values, for a large class of potentials
including the physically most interesting case of $N$-particle Coulomb
interaction with arbitrary charges and masses. While we consider in this
paper only particles without spin, \BM\ can be naturally defined for
spinor-valued \wfs\ as well \cite{Bell,BohmIII,Holland}. We shall deal with
spin, including the motion in a magnetic field, in a subsequent work.

The quantity of central importance for our proof of these results---as well
as for the question of the \SA\ of the \ham ---turns out to be the quantum
flux $J^{\psi} (q,t) = (j^{\psi} ,|\psi |^2)$, a $(d+1)$-vector,
with $j^{\psi} = \vpsi |\psi|^{2}$ the quantum probability current. The
absolute value of the flux through any surface in $\Omega \times \R$
controls the probability that a trajectory crosses that surface. Surfaces
of interest for us are the boundaries of neighborhoods around all the
singular points for \BM. Loosely speaking, the importance of the quantum
flux flows from the following insight: ``If there is no absolute flux into
the singular points, the singular points are not reached.''

We remark that the quantum flux is, in fact, important for most
applications of quantum physics, as well as for the mathematics revolving
around the self-adjointness of Schr\"{o}dinger operators. Heuristically,
the ``right'' behavior of the quantum flux at the critical points ensures
\SA\ of the \ham ---i.e., conservation of probability. But suppose we ask,
probability of what? The usual answer---the probability of finding a
particle in a certain region---is {\em justified}\/ by \BM : A particle is
found in a certain region because, in fact, it's there. By incorporating
the positions of the particles into the theo\-ry, and thus interpreting the
quantum flux as a flux of particles moving along trajectories, \BM\ can be
regarded as providing the basis for all intuitive reasoning in quantum
mechanics. (For more on this point, see also
\cite{Bohm52,BohmIII,Bell,DGZ1,DGZ3,Holland,DDGZ}.)

The paper is organized as follows: In Section 2, the relevant notion of
``typicality'' is discussed. Section 3 contains our main results. In
Section 3.1 we present the broad structure of the argument and in Section
3.2 we show how to transform the problem to that of controlling flux
integrals. The main theorem and corollary are proven in Section 3.3. In
Section 4 we discuss various aspects of the self-adjointness of the
Hamiltonian from the point of view of \BM . In particular, in Section 4.1
we show that in $d=1$ dimensions global existence holds under conditions
which in certain respects are milder than those of Theorem \ref{mainth}.

\bigskip

This is the first work concerned with a rigorous examination of the problem
of existence of the motion in \BM . For the related theory of Nelson,
stochastic mechanics, this question has been discussed by Nelson
\cite{Nelson1} and also by Carlen \cite{Carlen}. The behavior of the
Bohmian motion at the nodes of $\psi$ has been addressed by Bohm
\cite{Bohm52} and Holland \cite{Holland}. Bohm argues that particles are
either repelled from the nodes or cross them with infinite speed. (Bohm,
however, was not concerned with the question of existence but with
consistency with ``$\rho = |\psi |^2$.'') Holland claims to show that a
trajectory cannot reach a node unless it is always at some node. His
argument, however,  is circular, in that it requires the very regularity
whose breakdown at nodes is the source of difficulty.

Here is a simple counterexample to the claims of Bohm and Holland: Consider
the one-dimensional harmonic oscillator (with $\hbar=m=\omega=1$) and take
as the \wf\ of the particle a superposition of the ground state and the
second excited state, $\psi _t(q)= e^{-q^2/2} e^{-it/2} [ 1+(1-2q^2)
e^{-2it}]$. This \wf\ has nodes (among others) at $q=0$,
$t=(n+\frac12)\pi$ for all integers $n$. It leads to a velocity field which
is an odd function of $q$, i.e., which defines a motion which is reflection
invariant. Therefore $Q_t=0$, $t \neq (n+\frac12)\pi$, is a solution of
(\ref{bohmev}) which runs---first---into the node $(0,\pi/2)$ (with velocity
0 and which furthermore may be consistently continued through the nodes).

\section{Equivariance and Typicality}

The dynamical system defined by \BM\ is associated with a natural measure,
given by the density $|\psi_0 |^2$ on configuration space $\Omega$. If
$\psi_0$ is normalized, i.e., if the $L^2$-norm $\| \psi_0 \| = (\int
_{\Omega} |\psi_0 |^2 dq )^{1/2} =1$, then the density $|\psi_0 |^2$
defines a probability measure on configuration space $\Omega$, which we
shall denote by $\P$, that plays the role usually played by the
``equilibrium measure.'' Thus $\P$ defines our notion of ``typicality''
\cite{DGZ1}.  {\em Given}\/ the existence of the dynamics $Q_t$ for
configurations---the result we establish here---the notion of typicality is
time independent by equivariance \cite{DGZ1}:
\begin{equation} \label{equivar}
\rho_0=|\psi_0|^2 \Longrightarrow \rho_t = |\psi_t|^2 \ \mbox{ for all }\ t\in
\R ,
\end{equation}
where $\rho_t$ denotes the probability density on configuration
space $\Omega$ at time $t$---the image density of $\rho_0$ under
the motion $Q_t$. This follows from comparing the continuity equation for an
ensemble density $\rho _t(q)$
\begin{equation}
\frac{\partial \rho _t(q)}{\partial t} + \sum_{k=1}^{N}{\nabla}_{k} \cdot
[ {\bf v}^{\psi _t}_{k}(q) \rho _t(q)] =0  \label{conteq}
\end{equation}
with the quantum continuity equation
\begin{equation}
\frac{\partial |\psi_t(q)|^{2}}{\partial t} + \sum_{k=1}^{N}
{\nabla}_{k}\cdot {\bf j}^{\psi_t}_{k}(q)  =0 \label{qfluxeq}
\end{equation}
and noting that the quantum probability current
 $j^{\psi}=({\bf j}_{1}^{\psi},...,{\bf j}_{N}^{\psi})$ is given by
\begin{equation}
{\bf j}_{k}^{\psi}={\bf v}_{k}^{\psi}|\psi|^{2}=
\frac{\hbar}{{2im}_{k}} ({\psi}^{\ast}  {\nabla}_{k} \psi  -
\psi {\nabla}_{k} {\psi}^{\ast} ).  \label{qflux}
\end{equation}
We further denote the space-time current, the quantum flux, by  $J^{\psi} =
(j^{\psi}, |\psi |^2)$. In our proof of
global existence, this quantity gives the basic estimate for the probability
that a trajectory reaches singularities of
the velocity field or infinity.

It is at this stage important to bear in mind the conceptual difference
between the Equations (\ref{conteq}) and (\ref{qfluxeq}). The continuity
equation (\ref{conteq}), even without
global existence of differentiable trajectories $Q_t$, holds ``locally'' on the
set where $\vpsi$ is smooth, with
$\rho_t$ suitably
interpreted. This understanding is indeed basic to all our proofs.

Equation (\ref{qfluxeq}), on the other hand, is an identity for every $\psi_t$
which satisfies \seq\ classically. This
is seen by calculating
\begin{equation}
\frac{\partial |\psi_t|^{2}}{\partial t} =
\frac{1}{i\hbar}
({\psi}^{\ast} _t H \psi _t  -
\psi _t H {\psi}^{\ast} _t)  . \label{fluxH}
\end{equation}
But, without having established global existence, it is not a continuity
equation in the classical sense---despite its
name. By establishing global existence, we simultaneously show that the quantum
probability current $j^{\psi}$
is indeed a classical probability current, propagating the ensemble density
$|{\psi}|^{2}$ along the integral
curves of the vector field $\vpsi$.

\section{Global existence and uniqueness}

We make the following general assumptions:
\begin{deflist}{{\bf A8}:}
\item[{\bf A1}:] {\it The potential $V$ is a
$C^\infty$-function on $\Omega$.}
\item[{\bf A2}:] {\it The Hamiltonian $H$ is a self-adjoint
extension of $H|_{C_0^{\infty}(\Omega )}$ with  domain ${\cal D}(H)$.}
\item[{\bf A3}:] {\it The initial \wf\ $\psi_0$ is a $C^\infty$-vector of
$H$, $\psi_0
\in C^\infty (H)$, and is normalized, $\| \psi _0\| =1$.}
\end{deflist}
The boundary $\partial \Omega$ of the configuration space $\Omega$ will be
denoted by ${\cal S}$. (Recall that usually ${\cal S}$ is the set of
singularities of the potential.) $C_0^{\infty}
(\Omega )$, the set of
$C^\infty$-functions
with compact support contained in $\Omega$, is dense in $L^2(\Omega )$, and the
Hamiltonian is symmetric on
this set.
Since $H$ is real, i.e., commutes with complex conjugation,
there always exist self-adjoint extensions. The set of admissible initial \wfs
, $C^\infty (H) = \bigcap
_{n=1}^\infty {\cal D}(H^n)$, is dense in
$L^2(\Omega )$ and invariant under the time evolution $e^{-itH/
\hbar }$, and is therefore a core, i.e., a domain of essential
self-adjointness for $H$.\footnote{Some special $C^\infty$-vectors
are eigenfunctions and ``wave packets'' $\psi \in
\mbox{Ran}({P}_{[a,b]})$,
where $P_{[a,b]}$ denotes the spectral projection of $H$ to the
finite energy
interval $[a,b]$.}

In Lemma \ref{regl} we show that as a consequence of A1--A3 we may regard
$\psi_t = e^{-itH/\hbar } \psi_0 $
as being in
$C^{\infty }(\Omega \times \R)$ (and thus as
a classical solution of Schr\"odinger's equation).
Then the velocity field $\vpsi$ (cf.\ (\ref{velfield})) is
$C^\infty$ on the complement of the set ${\cal N}$ of nodes of
$\psi$, ${\cal N} := \{ (q,t) \in \Omega \times \R : \psi (q,t)=0
\}$, i.e., on the set of ``good'' points \[ {\cal G} := (\Omega \times \R )
\setminus {\cal N} ,\] which is an open subset of $\conf \times \R$. Let ${\cal
G}_t$ denote the slice of $\cal G$
at a fixed time $t$: ${\cal G}_t:= \Omega  \setminus {\cal N} _t $, where
${\cal N}_t:= \{ q \in \Omega : \psi
(q,t)=0\}$.
Then by a standard theorem of existence and uniqueness of ordinary
differential equations, for all initial values $(q_0,t_0)$ in ${\cal G}$ there
exist $\tau ^- (q_0,t_0) <t_0 $, $\tau
(q_0,t_0) >t_0$, and a unique maximal (non-extendible) solution $Q$
of (\ref{bohmev}) on the time interval $(\tau ^- (q_0,t_0), \tau
(q_0,t_0))$. {}From continuous dependence on initial values, the domain $D$ of
the maximal solution
$Q(t;q_0,t_0)$,
\be{defD} D:= \{ (t,q_0,t_0): (q_0,t_0)\in \g ,t\in (\tau ^- (q_0,t_0), \tau
(q_0,t_0)) \}, \ee
is an open subset of $\R ^{d+2}$ (and $Q$ is locally Lipschitz continuous
on $D$ with respect to $(t,q_0,t_0)$). Thus $\tau $ is lower
semi-continuous and hence, in particular, measurable. Because of the time
translation invariance of the theory, we may fix $t_0=0$, writing
$\tau(q_0)$ for $\tau (q_0,0)$, with similar notation for $\tau ^-$. Under
additional conditions on $\Omega$ and $H$ (see Corollary \ref{maincor}), we
shall show that $\tau (q_0)=\infty$ for typical $q_0$, i.e., we show that
the solution exists globally in time $\P$-almost surely:
\be{glo} \P (\tau < \infty ) =0. \ee This is equivalent to
\be{globT} \forall T<\infty: \ {\bf P}(\tau  < T )=0.      \ee
(Note that by time translation invariance and equivariance, $\P (\tau
<\infty )=0$ for all $\psi_0\in C^\infty(H)$ implies that $\P (\tau ^-
>-\infty)=0$ for all $\psi_0\in C^\infty(H)$, so that (\ref{glo}) indeed
implies global existence and uniqueness.)

\subsection{The program} \label{program}

We view the maximal solution $Q_t$ as a stochastic process on
${\cal G}_0$ equipped with the probability measure ${\bf P}$, i.e., $q_0$ is
distributed according to the
probability density $|\psi _0|^2$.
The basic criterion for global existence arises from the following properties
of the maximal solution. The set of
limit points $L(q_0)$ of the trajectory starting at $q_0$ ($q^* \in L $
:$\Leftrightarrow$ there is a sequence $t_k,
t_k \to \tau\equiv\tau(q_0)$ with $\lim _{k\to \infty} Q_{t_k} = q^*$) is
either empty---this is equivalent to $\lim
_{t\nearrow \tau} |Q_t| = \infty$---or nonempty, in which case, if $\tau <
\infty$, $(q^* ,\tau)\in \partial {\cal G}$
for all $q^* \in L$.
(The solution $Q$ need not be continuous at $t=\tau$, i.e., $L$
might contain more than one point, and there might
additionally be sequences $t_k \rightarrow \tau$ along which
$|Q_{t_k}| \rightarrow \infty$.) We thus have to see whether
trajectories  come too close to the boundary of $\g$ or to infinity. We do
this by checking whether they reach the boundary of $\g ^n$, an increasing
($\g ^{n_1} \subset \g ^{n_2}$ for $n_1 < n_2$) sequence of open sets $\g
^n \subset \g $, $\overline {\g ^n} \subset \g$, which are bounded in
configuration space, i.e.,  for all $T\in \R$,
the set $\g ^n _{[0,T]} := \g ^n \cap (\R ^d \times [0,T])$  is bounded.

For $q_0 \in \g ^n _0$, we introduce the stopping time $\tau ^{n}
(q_0)$, at which the process $Q_t$ first hits the boundary of $\g ^n$:
\[ \tau ^n (q_0) := \sup \{ s>0: (Q_t (q_0),t) \in \g ^n  \mbox{
for all } t\leq s\} . \]
Now, from the elementary theory of ordinary differential equations,
for all $q_0 \in \g ^n_0$,
\begin{equation} \label{rand} \mbox{if } \  \tau (q_0)<\infty ,
\mbox { then } \ \tau ^n(q_0) < \tau (q_0) \ \mbox{ and }
(Q_{\tau ^n(q_0)}(q_0),{\tau ^n (q_0)}) \in \partial \g ^n . \end{equation}
Furthermore, the sequence $\tau ^n$ is increasing in $n$.
For all $n$ and all $T\leq \infty $, we have
\[ \{ q_0 \in \g _0 :\tau < T \} \subset (\g _0 \setminus \g _0^n)
\cup \{ q_0 \in \g _0^n : \tau ^n < T \} \]
and therefore
\be{glob} \P ( \{ q_0 \in \g _0 :\tau < T \} ) \leq \P (\g _0 \setminus
\g _0^n)
+ \P( \{ q_0 \in \g _0^n : \tau ^n < T \} ) \ee
Thus to obtain the global existence and uniqueness of \BM\ for typical
initial configurations, it is sufficient to establish the vanishing of the
right hand side of (\ref{glob}) as $n\to\infty$ for some sequence of sets
$\g ^n$.  (Note as a matter of fact that the right hand side of
(\ref{glob}) decreases as $n$ increases.)

To proceed, we need to separate different parts of the boundary of $\g ^n$
which we shall treat in different ways: those close to infinity, those
close to ${\cal S}=\partial
\Omega $, and those close to the set $\cal N$ of nodes of the wave function. We
introduce ${\cal K}^{n}$, a sequence of bounded open sets exhausting
$\conf$, ${\cal K}^{n} \nearrow \conf$; ${\cal S}^\delta$, a sequence of
closed neighborhoods of $\cal S$ of ``thickness $\delta$''; and ${\cal
N}^\epsilon $, a sequence of closed neighborhoods of ${\cal N}$ of
``thickness $\epsilon$.'' (For the following general remarks, we do not
need to specify these sets more concretely; this will be done in Section
\ref{theorems}.) Thus the index $n$ accordingly gets replaced by
$\epsilon \delta n$. ${\cal G} ^{\epsilon \delta n}$ then denotes the set
of ``$\epsilon$-$\delta$-$n$-good'' points in configuration-space-time:
\[ {\cal G} ^{\epsilon \delta n}:=
((({\cal K} ^n\cap \Omega) \setminus {\cal S} ^\delta )
 \times \R ) \setminus {\cal N} ^\epsilon ,\]  and $ {\cal G} ^{\epsilon \delta
n}_t$ denotes the slice at a fixed
time $t\in \R$: \[ {\cal G} ^{\epsilon \delta n}_t:= ({\cal K} ^n\cap \Omega )
\setminus ({\cal S} ^\delta \cup
{\cal N} _t^\epsilon  ).\]
Furthermore, we define
\be{gdnTdef} \gdnt := (({\cal K}^n\cap \Omega ) \setminus {\cal S}^\d ) \times
(0,T). \ee  {}From (\ref{rand}), we
may write,
with $ x:= (Q_{\min (\tedl ,T)},\min (\tedl ,T))$,
\begin{eqnarray}  \{ q_0 \in \gedl _0: \tedl <T\} & = &  \{ q_0 \in \gedl _0: x
\in \partial \gedl \cap
(\conf
\times (0,T) ) \} \nonumber\\
& = &  \{ q_0 \in \gedl _0: x\in \partial {\cal N}^\epsilon \cap \gdnt \}
\nonumber\\
&  \cup &  \{ q_0 \in \gedl _0: x\in ( \partial {\cal S}^\delta \cap \Omega )
\times (0,T) \} \nonumber
 \\ &  \cup &  \{ q_0 \in \gedl _0: x\in (\partial {\cal K}^n  \cap \Omega
)\times
(0,T) \} .\label{split}
\end{eqnarray}
and therefore we arrive at
\begin{eqnarray}
\lefteqn{\P ( \{ q_0 \in \gedl _0:\tedl <T\}) \ \leq \
\P (  \{ q_0 \in \gedl _0:x\in \partial {\cal N}^\epsilon \cap \gdnt \} )
\hspace{6cm} \nonumber }\\
& & \hspace{2.5cm} + \ \P ( \{ q_0 \in \gedl _0: x\in (\partial {\cal S}^\delta
\cap \Omega ) \times (0,T) \} )
\nonumber\\
& &  \hspace{2.5cm} + \ \P ( \{ q_0 \in \gedl _0: x\in (\partial {\cal K}^n
\cap \Omega ) \times (0,T) \} )
 .\label{globc}    \end{eqnarray}
By virtue of (\ref{glob}) (almost sure) global existence follows if for {\it
some suitable}\/ choice of sets
${\cal N}^\epsilon$, ${\cal S}^\delta $, and ${\cal K} ^n$,  $\P (\g _0
\setminus \gedl _0)$ and the right hand side of (\ref{globc}) can be made
arbitrarily small by appropriately choosing $\epsilon$, $\delta$ and $n$.

\subsection{The flux argument}\label{fluxargument}

Consider the random trajectory $(\g _0, \P , \wt Q_t)$ obtained by
stopping the original process $Q_t$ at time $\tau$ and placing it in the
cemetery $\dag$\ :
 The process $\wt Q_t: \g _0
 \longrightarrow
\Omega \cup \{ \dag \} $ is defined,  for all $t\geq 0$, by
\begin{equation}
\wt Q_t (q_0):= \left\{\begin{array}{cl}
Q_t(q_0)     & \mbox {for } t< \tau (q_0) \\
\dag       & \mbox {for } t \geq \tau (q_0).
\end{array}
\right.           \label{process}
\end{equation}
Let $\rho_t$ be the image density  of  $\wt Q_t$ restricted to $\Omega$.

Denote by
${\cal I}$ the set \[ {\cal I} := \{ (Q_t (q_0), t) :  t\in (\tau ^-(q_0), \tau
(q_0)) \ \mbox{and}
\ q_0 \in \g _0\} ,\]and by   ${\cal I}_t:= \mbox{Ran} \wt Q_t \setminus \{
\dag \}$ (${\cal
I}_t
\subset \g _t$) its time-$t$ slice.
$ {\cal I}$ is an open subset of $ \g$. ($\cal I$ can be identified with $D
\cap (\{ 0\} \times \R ^{d+1})$, cf.\ (\ref{defD}).) Clearly $\rho _t = 0$ on
$\g_t \setminus {\cal I}_t$ for $t>0$.
Note that on $\cal I$ both $|\psi _t|^{2}$ and $\rho _t$
are solutions of the continuity equation (\ref{conteq}) restricted
to $\cal I$ with the same initial data. Uniqueness of solutions of quasilinear
first order
partial differential equations on the set where the characteristics exist
implies that for all $t\geq0$
\begin{equation}
\rho _t (q) = |\psi _t(q)|^{2} \;\; \mbox{for all}\,  q\in {\cal I}_t.
\label{rhobound}
\end{equation}

Consider now a smooth surface $\Sigma$ in $\g$. Recalling the
probabilistic meaning of the flux $ J_t(q):= (\rho_t(q)
\vpsit (q), \rho _t(q))$, we obtain that
the expected number of crossings of $\Sigma$  by the random trajectory $\wt
Q_t$ (including tangential
``crossings'' in which the the trajectory remains on the same side of
$\Sigma$) is given by
\be{afi} \int _{\Sigma}|J_t(q) \cdot U| d\sigma  \ee
where $U$ denotes the local unit normal vector at $(q,t)$.
($\int_{\Sigma}(J \cdot U)d\sigma $ is the expected number of {\em signed}\/
crossings.)
(Consider first a small surface element which the trajectory can cross at
most once. The probability density for this crossing is readily calculated
to be $|J \cdot U|$. Invoking the linearity of the expectation value
yields then the general statement.)\footnote{In stochastic mechanics
\cite{Nelson1}, which involves the same quantum flux, the particle
trajectories are realizations of a diffusion process and are hence not
differentiable, i.e., velocities do not exist. Thus in stochastic mechanics
the flux does not have the same probabilistic significance and hence the
subsequent arguments are not valid for stochastic mechanics.}

The probability of crossing $\Sigma $ (at least once) is hence bounded by
(\ref{afi}). {}From (\ref{rhobound}) we obtain that
\[ |J_t \cdot U| \leq |(|\psi _t(q)|^2 \vpsit (q), |\psi_t|^2)
\cdot U| = |(j^{\psi_t} , |\psi _t |^2) \cdot U| = |J^{\psi _t} \cdot U| \]
and thus we arrive at the bound
\be{crossb} {\bf P}( \wt Q_t \mbox{ crosses } \Sigma ) \leq \int _{\Sigma }
|J^{\psi_t} (q) \cdot U| d\sigma .\ee
If now the sets ${\cal N}^\epsilon $, ${\cal S}^\delta $, and ${\cal K}^n$ are
choosen in such a way that their
boundaries are piecewise integrable surfaces,
the events on the r.h.s.\ of  (\ref{globc}) are crossings by $\wt Q_t$ through
the
respective surfaces, and hence (\ref{crossb}) implies the following
bounds for the  terms in  (\ref{globc}):
\begin{eqnarray}
 \P ( x\in (\partial {\cal N}^\epsilon \cap \gdnt )) & \leq & \int
_{\partial {\cal N}^\epsilon \cap {\gdnt}} |J^{\psi_t} (q) \cdot U| d\sigma
:= {\bf N}(\e ,\d ,n) ,\nonumber\\ \P ( x\in ((\partial {\cal
S}^\delta\cap\Omega) \times (0,T)) ) & \leq & \int _{(\partial {\cal
S}^\delta\cap\Omega) \times (0,T)} |J^{\psi_t} (q) \cdot U| d\sigma := {\bf
S}(\d ) , \label{fluxest}\\ \P ( x\in ((\partial {\cal K}^n \cap \Omega )
\times (0,T) ) ) & \leq & \int _{(\partial {\cal K}^n \cap {\Omega})\times
(0,T)} |J^{\psi_t} (q) \cdot U| d\sigma =: {\bf I}(n)
.\nonumber\end{eqnarray} (If a boundary happens to be the empty set, the
corresponding integral of course vanishes.)

It seems intuitively rather clear\footnote{By mentioning these heuristics
we do not wish to suggest the structure of the rigorous proof given in the
next section, nor need this proof sustain these heuristics.} that all the flux
integrals should
vanish in the limit $\epsilon \to 0,\delta \to 0 $, and $n\to \infty$: It
seems fairly obvious that  the ``nodal integral'' ${\bf N}(\e ,\d ,n)$
should vanish as $\epsilon \rightarrow 0$ since $J^{\psi _t}$ is zero at the
nodes.\footnote{One might worry
about the
``size of $\partial \No$''being uncontrollably large. However,
since $\psi$ is a complex smooth function,
${\cal N}$ might be expected to  have codimension 2 ``generically,'' so
$\partial {\cal N}^\epsilon $ should have small area.}
The ``singularity integral'' ${\bf S}(\d )$ should vanish
in the limit $\delta \to 0$ if the set $\cal S$ has codimension greater
than 1, which is usually the case. Furthermore, $j^\psi =0$ at $\cal S$ is a
natural boundary condition defining a
domain of \SA\ of the Hamiltonian. Finally, the ``infinity integral'' ${\bf
I}(n)$ should tend to zero as $n \to
\infty$ since $\psi _t(q)$ (which is sufficiently smooth) and hence $J^{\psi_t}
(q)$ should
rapidly go to zero as $|q| \to \infty$.

\subsection{Global existence of \BM } \label{theorems}

Our main result is the following theorem:
\begin{theorem} \label{mainth}
 Assume {\rm A2, A3,} and further
\begin{deflist}{{\bf A8$^\prime$}{\rm :}}
\item[{\bf A1$^\prime$}{\rm :}] {\rm A1} and ${\cal S} \subset  \bigcup
_{l=1}^m {\cal
S}_l$, where $m <\infty$ and the ${\cal S}_l$ are $(d-3)$-dimensional
hyperplanes;
\item[{\bf A4}{\rm :}]  $\int _0^T \| \nabla \psi _t\| ^2 dt < \infty $
for all \ $0<T<\infty$.
\end{deflist}
 Then $\P (\tau <\infty )=0$.
\end{theorem}

 Since ${\cal S}_l$ is a $(d-3)$-dimensional
hyperplane, it may be written as ${\cal S}_l =\{ {\bf y}_l ={\bf a}_l\} $
with ${\bf
y}_l$ denoting the map $\R ^d \to \R ^3, q\mapsto (q\cdot y_l^1, q\cdot
y_l^2, q\cdot y_l^3)$ where $y^1_l, y^2_l, y^3_l$ are 3 orthogonal unit
vectors normal to the hyperplane ${\cal S}_l$ and ${\bf a}_l\in \R ^3$ a
constant.

The Condition A1$^\prime$ on the shape of $\cal S$ fits well with the
3-dimensionality of physical space.  If  $V$ is a central potential, ${\cal
S}_l$ is of the form $\{ {\bf q}_i =0\}$, and for a pair potential, ${\cal
S}_l$ is of the form $\{ {\bf q}_i-{\bf q}_j =0 \}$ for some $1\leq i<j\leq N$.
(Note that if $d=\nu N<3$,
Assumption A1$^\prime$ demands that ${\cal S}=\emptyset$.)

Under the Assumption A1$^\prime$, the configuration space  $\Omega = \R
^d\setminus {\cal S}$; in particular, $L^2(\Omega )=L^2(\conf )$.\footnote{Thus
Theorem \ref{mainth} does
not cover the case of a bounded configuration space $\Omega$, for which
boundary conditions of Dirichlet or
Neumann (or mixed) type are normally imposed. See, however, our Theorem
\ref{d1theor}.} Recall that $H_0$
denotes the \sa\ operator
\[ H_0 = - \sum _{k=1}^N \frac{\hbar ^2}{2m_k} \Delta _k\]
on the Hilbert space ${\cal H}= L^2(\Omega )=L^2(\conf )$.

The  Condition A4 of ``finite integrated kinetic energy''
may be ensured by bounding the quadratic form
$(\nabla\psi_t,\nabla\psi_t) \leq
M (\psi_t, H_0 \psi_t)$ with $M = (2/ \hbar ^2) \max (m_1,\dots ,
m_N)$ by the form $(\psi_t, H \psi_t)$, which is finite and
independent of $t$ for $\psi_0$ (and hence $\psi_t$) in the form
domain \cite{RSI} ${\cal Q}(H) (\supset {\cal D}(H))$ of the Hamiltonian
$H$.\footnote{Note that the notation $(\psi ,A\psi )$ for the quadratic
form associated with the self-adjoint operator $A$ is symbolic: Only for
$\psi \in {\cal D}(A)$ does it coincides with the indicated scalar product
in ${\cal H} = L^2 (\conf )$; more generally it can be defined via the
spectral representation for $A$.}
The following corollary shows that Theorem \ref{mainth} indeed implies the
global
existence and uniqueness of Bohmian mechanics for all $\psi_0\in
C^\infty(H)$  for a large class of Hamiltonians.
\begin{coroll} \label{maincor} Assume
\begin{deflist}{{\bf A8$^{\prime \prime}$}{\rm :}}
\item[{\bf A1$^{\prime \prime}$}{\rm :}] {\rm A1$^{\prime}$} and $V=V_1+V_2$,
where
$V_1$ is bounded below, and $V_2$ is $H_0$-form bounded with relative bound
$a<1$,
\item[{\bf A2$^\prime$}{\rm :}] $H$ is the form sum $H_0+V$ \cite{Faris},
\end{deflist} and {\rm A3}. Then $\P (\tau <\infty )=0$ and \BM\ exists
uniquely and
globally in time $\P$-almost surely.
\end{coroll}

\proof We show that A4 holds:   That
$V_2$ is $H_0$-form bounded means that ${\cal Q}(H_{0})\subset {\cal
Q}(V_2)$ and that for $\psi \in {\cal Q}(H_0)$ there exist constants $a,b >0$
such that \[
|(\psi, V_2 \psi )|\leq a(\psi , H_{0}\psi ) + b (\psi ,\psi ). \]
Since $V_1(q)\geq -c,\, c>0$, for  all
$q\in \Omega$, we obtain for $\psi \in {\cal Q}(H) = {\cal Q}(H_0) \cap {\cal
Q}(V_1)$ that
\begin{eqnarray*} (1-a) (\psi , H_0 \psi )& \leq &
(\psi , (H_0+V_2) \psi ) + b(\psi , \psi) \\
& \leq & (\psi , (H_0 + V_1 + V_2) \psi ) + c(\psi ,\psi ) + b(\psi ,\psi )\\
&=& (\psi ,H\psi )+(b+c) (\psi ,\psi ) .\end{eqnarray*}
Hence with $a<1$ we have that for $\psi_0 \in {\cal Q}(H) \subset {\cal
Q}(H_0)$ and all $t$
 \begin{eqnarray*} \frac 1M (\nabla \psi _t,\nabla\psi _t) \leq
(\psi _t, H_{0} \psi _t) \leq
\frac 1{1-a}  (\psi _t, H \psi
_t) + \frac{b+c}{1-a} (\psi_t ,\psi_t) & & \\
= \frac 1{1-a}  (\psi _0, H \psi
_0) + \frac{b+c}{1-a} \| \psi_0 \| ^2 & & \end{eqnarray*}
and A4 follows. \eop

\bigskip

The class of $H_0$-form bounded potentials, with arbitrary small relative
bound $a$, includes for example $R+L^\infty $ or $L^{3/2}+L^\infty$ on $\R^3$,
where $R$ is the Rollnik
class. (For details, see for example \cite{Katob,Simon,ReeSiII}.) Therefore
such $H_0$-form bounded potentials
include power law interactions $1/r^\alpha $ with $\alpha <2$, and thus the
physically most relevant potential of
$N$-particle Coulomb interaction with arbitrary charges and masses. (The class
of $H_0$-form bounded
potentials contains the more familiar class of $H_0$(-operator) bounded
potentials, which already includes the
$N$-particle Coulomb interaction \cite{kato51}.)
Furthermore, harmonic and anharmonic (positive) potentials are included, and
arbitrarily strong positive
repulsive potentials.

\bigskip
\noindent
{\it Proof of Theorem \ref{mainth}.} \quad
We establish (\ref{globT})---for all $0<T<\infty$, $ \P (\tau <T)
=0$---following the program  described in
Section \ref{program}
and the flux argument of Section \ref{fluxargument}.

We first choose suitable sets
${\cal N}^\epsilon$, ${\cal S}^\delta $, and ${\cal K} ^n$. Let $\e >0$. Set
\be{nepsdef} {\cal N}^\epsilon := \bigcup _{ k  :C^\e (k) \cap
{\cal N}
\neq \emptyset  } C^\e (k),\ee
where $(C^\e (k))_{k \in \N }$ is a ``partition'' of
configuration-space-time into closed cubes with side length $\e$ whose
edges are parallel to the canonical basis vectors of $\R ^{d+1}$. Let $ \d =
(\d _1, \dots , \d
_m)$, $\delta _l> 0$ for all $l$. Recalling that ${\cal S} \subset \bigcup
_{l=1}^m {\cal S}_l$ with ${\cal S}_l = \{ {\bf y}_l ={\bf a}_l\}$), set
\[ {\cal S} ^{\delta} := \bigcup _{l=1}^m {\cal S}
_l^{\delta _l},
\quad {\cal S} _l^{\delta _l} := \{ q\in \conf : \mbox{dist}
(q,{\cal S} _l)
\leq \d _l \} = \{ |{\bf y}_l-{\bf a}_l| \leq \d _l\} .\]
For the cutoff at infinity we choose open balls with radii $n\in \R ^+$:
\[ {\cal K} ^n  :=  \{ q\in \conf : |q|<n\} . \]
By virtue of (\ref{glob}), (\ref{globc}), and (\ref{fluxest}), we obtain that
for all $0<T<\infty$
\begin{eqnarray} \P (\tau < T ) & \leq & \P (\g _0 \setminus \gedl
_0)
+ \P (\tedl <T) \nonumber\\
& \leq &  \P (\g _0 \setminus \gedl _0) +
\P ( x\in (\partial {\cal N}^\epsilon \cap \gdnt )) \nonumber\\
& & + \P ( x\in (\partial {\cal S}^\delta  \times (0,T)) )
 + \P ( x\in ((\partial {\cal K}^n  \cap \Omega )\times (0,T) ))
\nonumber\\
& \leq &  \P (\g _0 \setminus \gedl _0) +
{\bf N}(\e ,\d ,n) + {\bf S}(\d ) + {\bf I}(n) \label{SD}\end{eqnarray}

For the first term on the right hand side of (\ref{SD}) recall that $\gedl _0 =
({\cal K} ^n \cap \Omega ) \setminus
( {\cal N}_0^{\epsilon} \cup {\cal S} ^{ \delta})  $; therefore
\[ \g _0 \setminus \gedl _0 = ( \g _0 \setminus {\cal K} ^n ) \cup
(\g _0 \cap {\cal S} ^{ \delta}) \cup (\g _0 \cap {\cal
N}_0^{\epsilon} ),\]
and thus
\[ \P (\g _0 \setminus \gedl _0 )\leq \P ( \g _0 \setminus {\cal K} ^n ) + \P
(\g _0 \cap {\cal S} ^{ \delta}) +\P (\g
_0 \cap {\cal N}_0^{\epsilon} ).\]
The vanishing of the three terms on the right hand side
in the limit $n\to \infty$, $\d \to 0$, resp.\ $\e \to 0$, follows easily from
the facts that $\P$ is a probability
measure with density  $ {|\psi _0|}^2$, and that the respective sets tend to
$\P$-measure 0 sets.

The vanishing of the remaining terms in (\ref{SD}) is the content of the
following lemmas:

\begin{lemma} \label{infl} Assume {\rm A1--A4}. For all
$0<T<\infty$
there exists a sequence $n_k$, $n_k \to \infty$ as
$k\to \infty$, with
\[ \lim _{k\to \infty} {\bf I}(n_k) =0.\]
\end{lemma}

\begin{lemma} \label{singl} Assume {\rm A1$^\prime$, and A2--A4}.
Then there exists a sequence of $m$-vectors $ \delta^{(k)}$, $|\delta^{(k)}|
\to 0$ as $k\to \infty$ (with
$\delta_l^{(k)} >0$ for all $l,k$), with
\[ \lim _{k\to \infty} {\bf S}( \d^{(k)}) =0.\]
\end{lemma}

\begin{lemma} \label{sdnodesl} Assume {\rm A1--A3}.
For all $0<T<\infty$,\ $n<\infty $ and $\delta >0$,
\[ \lim _{\epsilon \to 0} {\bf N}(\e ,\d ,n) =0.\]
\end{lemma}

These lemmas will be proven below.  Lemmas \ref{infl}, \ref{singl}, and
\ref{sdnodesl} imply that the r.h.s.\ of (\ref{SD}) can be made arbitrarily
small. (Note that if $d<3$, Assumption A1$^\prime$ demands that ${\cal S} =
\emptyset$, and hence that ${\bf S}\equiv 0$, so that Lemma  \ref{singl} is
trivial in this case.)
\eop

\bigskip
\noindent
{\it Proof of Lemma  \ref{infl}.} \quad
\begin{eqnarray*}
{\bf I}(n) & = & \int _{(\partial {\cal K}^n \cap {\Omega})\times
(0,T) }
|J^{\psi_t} (q) \cdot U| \ d\sigma
\  = \  \int _0^T  \int _{\partial {\cal K}^n \cap \Omega
}|j^{\psi_t}
(q) \cdot u|\ ds\, dt \\
& \leq & \mu \int _0^T  \int _{\partial {\cal K}^n \cap \Omega
}|{\psi_t}| \
|\nabla \psi _t| \ ds \, dt \ =: \ \mu {\bf \tilde I}(n)
\end{eqnarray*}
with $\mu = \hbar / \min (m_1,\dots ,m_N)$, $ds$  the $(d-1)$-dimensional
surface element of
$\partial {\cal K}^{n}$, and $u$ the local unit normal
vector of this surface.
To show that ${\bf \tilde I}(n)$ goes to 0 along some sequence $n_k$, we prove
a stronger statement, namely
that ${\bf \tilde I}(n)$ is integrable over $n$. This is immediate since
$\dps \int _0^\infty {\bf \tilde I}(n)\, dn $ yields the
{\em volume}\/ integral of $|\psi_t |\ |\nabla \psi_t |$, which is
easily estimated:
\begin{eqnarray*}\int_0^\infty  {\bf \tilde I}(n)\, dn & = &
\int_0^T \int _\Omega    |{\psi_t}| \ |\nabla \psi _t| \ dq \, dt
 \\
& \leq & \int _0^T  \| \psi_t \| \, \| \nabla \psi _t \|  \ dt \
 =\   \int _0^T  \| \nabla \psi _t \|  \ dt \ <  \infty,
\end{eqnarray*}
where we have used A4 for the last inequality.
We may thus conclude that there exists a sequence $ (n_k)_k$ with $n_k \to
\infty$ as $k\to \infty$, along which
${\bf \tilde I}(n_k) \to 0$. This proves Lemma \ref{infl}. \eop

\bigskip
\noindent
{\it Proof of Lemma \ref{singl}.} \quad  We may assume that $d\geq3$.
We shall use the following Inequality: For $\psi \in {\cal Q}(H_0)$
\be{ineq} \int _{\conf} \frac{|\psi |^2}{4|{\bf y}_l-{\bf a}_l|^2} dq \leq
\int _{\conf } |\nabla \psi |^2 dq. \ee
This is a straightforward extension of the inequality
known as Hardy's inequality or the ``uncertainty principle lemma'' (see, for
example, \cite{ReeSiII})
usually given for $\psi \in C_0^\infty (\R ^3)$:
\[ \int _{\R ^3} \frac{|\psi |^2}{4r^2} d{\bf r} \leq
\int _{\R ^3} |\nabla \psi |^2 d{\bf r} . \]
(One immediately obtains (\ref{ineq}) for $d=3$ and $\psi\in C_0^\infty (\R
^3)$. Then, viewing $\psi \in C_0^\infty (\R ^d)$ as $\psi \in C_0^\infty
(\R ^3)$ by keeping all coordinates fixed except ${\bf y}_l$, one extends
this inequality easily to $C_0^\infty (\R ^d)$. It is then further extendible
to  $\psi \in {\cal Q}(H_0)$ because
$C_0^\infty (\R ^d)$ is dense in ${\cal Q}(H_0)$ with respect to the $H_0$-form
norm.)

First we estimate
\begin{eqnarray*}
{\bf S}( \d ) & = & \int _{\partial {\cal S}^{ \d} \times (0,T)} |J^{\psi_t}
(q) \cdot U| \ d\sigma \  = \  \int _0^T
\int _{\partial {\cal S}^{ \d } }|j^{\psi_t}
(q) \cdot u|\ ds\, dt \\
& \leq & \mu \sum _{l=1}^m \int _0^T  \int _{\partial {\cal
S}_l^{\d _l} \cap \Omega }
|{\psi_t}| \ | \nabla \psi _t| \ ds \, dt \ =: \ \mu \sum _{l=1}^m
\wt {\bf S}_l(\d _l).
\end{eqnarray*}
We now integrate $(1/|{\bf y}_l-{\bf a}_l|) \wt {\bf S}_l (\d _l )$
over $\d _l = |{\bf y}_l-{\bf a}_l|$: By the definition of ${\cal S} _l^{\delta
_l}  = \{ |{\bf y}_l-{\bf a}_l| \leq \d
_l\} $, this yields the
{\em volume}\/ integral of $ (|\psi |/|{\bf y}_l-{\bf a}_l|) \ |\nabla \psi |$,
 which may be bounded as follows:
\begin{eqnarray*}
& & \int_0^\infty \frac 1{\d _l} \wt {\bf S}_l(\d _l)\, d\d _l \ = \
\int_0^T \int _\Omega  \frac  {|{\psi_t}|}{|{\bf y}_l-{\bf a}_l|}  \ |\nabla
\psi _t| \ dq \, dt  \\
& & \hspace{1.5cm} = \ \int_0^T
\int _{\conf} \frac  {|{\psi_t}|}{|{\bf y}_l-{\bf a}_l|} \ |\nabla \psi _t|
\ dq \, dt
\ \leq \ \int _0^T  \| \frac {\psi_t }{|{\bf y}_l-{\bf a}_l|} \| \, \| \nabla
\psi _t \|  \ dt \\
&  & \hspace{1.5cm} \leq \ 2  \int _0^T  \| \nabla \psi _t \|^2  \ dt \ < \
\infty
\end{eqnarray*}
using Schwarz's inequality and the Inequality (\ref{ineq}).
Since $1/\d _l$ is not integrable at $\d _l=0$, for each $l$ there exists a
sequence $\d_l^{(k)}$ with $\d_l^{(k)}
\to 0$ as $k\to \infty$, along which $\wt {\bf S}_l(\d_l^{(k)}) \to 0$. This
proves
Lemma \ref{singl}.\eop

\bigskip
\noindent
{\it Proof of Lemma  \ref{sdnodesl}.} \quad
This proof is more involved than the previous ones, since the nodal set is
unknown. The basic idea is the following: Where the $(d+1)$-gradient $\dps
\psi ' = \Bigl( \nabla \psi, \frac{\p \psi }{ \p t}\Bigr) $ is small the
current is very small, and where $\psi '$ is not small the surface area can
be controlled.

Let $\eta >0$. We split the part of ${\cal N}^\e$ contributing to the surface
$\p {\cal N}^\e \cap \gdnt$ into two
(not necessarily disjoint) sets:
\[ {\cal N}^\e _{>} := \bigcup _{k \in I_> } C^\e (k), \quad
 \mbox{and} \quad
 {\cal N}^\e _{<} := \bigcup _{k \in I_< } C^\e (k) \quad
\mbox{with}\]
\[ I_> := \{ k:C^\e (k) \cap
\{ (q,t): \psi (q,t) =0, |\psi '(q,t)|>\eta \}
\cap \gdnt  \neq \emptyset \}  \]
\[ I_< := \{ k:C^\e (k) \cap
\{ (q,t): \psi (q,t) =0, |\psi '(q,t)| \leq \eta \}
\cap \gdnt  \neq \emptyset \} \]
Then
\begin{equation} \label{est1}
{\bf N}(\e ,\d ,n) = \int _{\partial {\cal N}^\e \cap \gdnt} |J^{\psi
_t}(q)\cdot U| \ d\sigma \ \leq \
\int _{\partial {\cal N}^\e_{>}} |J^{\psi _t}(q)| \ d\sigma +
\int _{\partial {\cal N}^\e_{<}} |J^{\psi _t}(q)| \ d\sigma
\end{equation}

On the compact set $\overline {\g ^{(\d /2) (n+1)}_{(-1,T+1)}}$ (cf.\
(\ref{gdnTdef})) there exist a global Lip\-
schitz constant $L$ for $\psi '$, and a global bound $K$ for $| \psi '|$.
Observe that for $\e < \min (\d /(2 \sqrt
{d}), 1/\sqrt{d})$, $\No ^\e  _{\stackrel ><} \subset \g ^{(\d /2)
(n+1)}_{(-1,T+1)}$. Let therefore  $\e < \min
(\d /(2\sqrt {d}), 1/ \sqrt{d})$.

\bigskip

Consider first ${\cal N}^\e_{<}$. In this set the flux $|J^\psi |$ is very
small. We may estimate the integral by simply taking an appropriate bound
of $|J^\psi |$ times the total area of the surfaces of {\em all}\/ the cubes.
In every $\e$-cube $C^\e$ of ${\cal
N}^\e_{<}$ there is a
point $(q^*, t^*)\in {\cal N}$ with $| \psi '(q^*,t^*) |\leq \eta$. Thus (in
every $\e$-cube of ${\cal N}^\e_{<}$
and hence) for all $(q,t)\in {\cal N}^\e_{<}$
\[ | \psi '(q,t)| \leq \eta + L \sqrt {d+1} \e . \]
$|\psi | $ is thus bounded on (every $\e$-cube of ${\cal N}^\e_{<}$ and
hence on) ${\cal N}^\e_{<}$ by $ (\eta +L \sqrt {d+1} \e )\sqrt {d+1} \e =:
c_1 \eta \e +c_2 \e ^2$. The flux is
then bounded by
\begin{eqnarray}
|J^\psi |=\sqrt {(|\psi |^2 )^2 + |j^\psi |^2 }& \leq & |\psi |^2+|j^\psi|
\leq |\psi |^2 + \mu |\psi | \, |\nabla \psi |  \label{jbound} \\ &\leq &
(c_1 \eta \e +c_2 \e ^2)^2 + \mu (c_1 \eta \e +c_2 \e ^2) (\eta
+L\sqrt{d+1} \e ) \nonumber \end{eqnarray} To bound the surface area of
${\cal N}^\e_{<} $, we simply add the areas of the surfaces of all $\e$-cubes
in $\g ^{(\d /2) (n+1)}_{(-
1,T+1)}$. The number of $\e$-cubes in $\g ^{(\d /2) (n+1)}_{(-1,T+1)}$ is
bounded by $c_3/\e ^{d+1}$ with \[
c_3(n,T,d)=(T+2)(2n+2)^d ,\] and the surface area of a single cube is
equal to $2(d+1) \e ^d$. Thus for the surface area of $\partial {\cal N}^\e
_<$ we have the bound
\be{check} |\partial {\cal N}^\e _< | \leq \frac {2(d+1)c_3}\e \ee
and combining (\ref{jbound}) and (\ref{check}) we obtain that
\begin{eqnarray}
& & \int _{\partial {\cal N}^\e_{<} }|J^\psi|d\sigma \leq \left(\sup
_{{\cal N}^\e_{<}} |J^{\psi} |\right) \,\left( |\partial {\cal
N}^\e_{<}|\right) \nonumber \\ & & \leq \frac {2(d+1)c_3}\e ((c_1 \eta \e
+c_2 \e ^2)^2 + \mu (c_1 \eta \e +c_2 \e ^2) (\eta +L\sqrt{d+1} \e ) )
\label{n-}\\ & & =O(\eta^2),\quad \e\to0.\nonumber \end{eqnarray}

\bigskip

Consider next the set ${\cal N}^\e_{>}$. On this set we can control the
size of the nodal surface. To do this we use a further partition of
configuration-space-time into cubes $(C^\gamma (k))_{k\in \N}$ of side length
$\gamma$ (with sides parallel to
the sides of the $C^\e$-cubes). We choose $\gamma$ so small that any
$C^\gamma$-cube which contains or overlaps the interior of a $C^\e$-cube of
${\cal N}^\e_{>}$ lies completely in $\g ^{(\d /2) (n+1)}_{(-1,T+1)}$,
i.e., $\gamma < \min (\d /(2\sqrt {d} )-\e ,(1/\sqrt {d}) -\e )$.
($\gamma $ will later be chosen to be proportional to $\eta$, and we  shall
take the limit $\e \to 0$ for fixed  $\eta$, so that $\gamma \gg \e$
eventually.)

We show now that in each $\gamma$-cube the number  of $\e$-cubes in ${\cal
N}^\e_{>}$ is small, at least compared with $ {( \frac{\gamma}{\e})}^{d+1}$.
Consider a $\gamma$-cube $C^\gamma (k)$ containing or
overlapping the interior of a $C^\e $-cube of  ${\cal N}^\e_{>}$. Then there is
a point $(q^*,
t^*)\in {\cal N}\cap C^\gamma_\e (k)$ with $| \psi '(q^*,t^*)|>\eta $, where
$C^\gamma_\e(k)$ is the ``$\e$-
fattened''
$\gamma$-cube, i.e., the cube of side $\gamma+2\e$ with the same center as
$C^\gamma (k)$. That $|\psi '(q^*,t^*)|>\eta $ implies that
\[ | \psi _i '(q^*,t^*)|>\frac \eta {\sqrt 2} \]
for either $ i= 1$ or $ i=2$ (or both), with $\psi _1 := \mbox {Re} \, \psi$
and $\psi _2 := \mbox {Im}\, \psi $.

Let $e_k$ be that basis vector which is closest to the direction of $\psi _i
'(q^*,t^*)$, i.e.,  for which $| e_k \cdot
 \psi _i '(q^*,t^*) |$  is maximal. Thus
\[ | e_k \cdot  \psi _i '(q^*,t^*) | > \frac \eta {\sqrt{2(d+1)}} \]
and hence we have that for all $(q,t) \in C^\gamma_\e (k)$
\[ |e_k \cdot \psi _i '(q,t) |> \frac \eta {\sqrt{2(d+1)}} -
L \sqrt {d+1} (\gamma + 2\e ). \]
Now choose $\gamma$ such that $L \sqrt {d+1} (\gamma + 2\e )  =
\eta /\left( 2\sqrt{2(d+1)}\right)$, i.e., introduce $c_4:= 1/\left( 2L \sqrt
{2}(d+1)\right)$ and set  $\gamma =
c_4\eta -2\e$. Then for all $(q,t) \in C^\gamma_\e (k)$
\be{gradpsii} |e_k \cdot \psi _i '(q,t) |> \frac
 \eta { 2\sqrt{2(d+1)}} . \ee

Let $x$ and $y$ be two space-time points in $C^\gamma_\e(k) \cap \No ^\e$ with
$\psi(y)=0$ and $x-y = l e_k, l>0$. Then,  on the one hand, by the global
bound $K$ on $| \psi '|$ we have that
\[ |\psi _i(x)| \leq  K\sqrt {d+1} \e. \] On the other hand,  it follows from
 (\ref{gradpsii}) that
$|\psi _i(x)| \geq l  \eta / (2\sqrt{2(d+1)})$. Thus  $l\leq  2K
\sqrt{2}(d+1)\e /{\eta}=: c_5\e / \eta $. Therefore the number of
$\e$-cubes in ${\cal N}^\e_{>}$ contained in $C^\gamma_\e(k)$ and lying in an
$e_k$-column---the set of
$e_k$-translates of an $\e$-cube---is bounded by $(c_5 / \eta) +1$. (This is a
rather crude estimate. The number of such cubes is in fact bounded by
$2d+\sqrt d +2$, independent of $\eta$, as can easily be seen by controlling
also the projection of $\psi_i'$ orthogonal to $e_k$.)

  Now the number of $e_k$-columns in $C^\gamma_\e(k)$ is
no greater than ${[(\gamma / \e ) +2]}^d$, while the
number of $\gamma$-cubes in $\g ^{(\d /2) (n+1)}_{(-1,T+1)}$ is bounded by $c_3
/\gamma ^{d+1}$.
Thus we obtain a bound for the surface area of $\No ^\e _{>}$:
\[ |\partial \No ^\e _{>} | \leq \left( \frac {c_5}{\eta }+1\right)
{\left(\frac {c_4\eta}{ \e }\right)}^{d} \frac {c_3 }{(c_4\eta -2\e )^{d+1}}
2(d+1) \e ^{d}. \]

$|J^\psi |$ may be estimated (as in
(\ref{jbound})) by invoking now the global bound $K$ for $| \psi '|$ to yield
\[ |J^\psi |\leq K^2 (d+1) \e ^2 + \mu
K^2 \sqrt {d+1} \e \]
on ${\cal N}^\e_{>}$. Thus we arrive at the estimate
\begin{eqnarray}
{\lefteqn {\int _{\partial {\cal N}^\e_{>} }|J|d\sigma }\label{n+}}\\ & &  \leq
\left( \frac {c_5}{\eta } +1\right)
{\left(\frac {c_4\eta}{ \e }\right)}^{d} \frac {c_3 }{(c_4\eta -2\e )^{d+1}}
2(d+1)
\e ^{d}  \Bigl( K^2 (d+1) \e ^2 + \mu K^2 \sqrt {d+1} \e \Bigr) \nonumber
\end{eqnarray}
Using (\ref{n-}) and (\ref{n+}), by letting first $\e \to 0$
and then $\eta \to 0$, it follows from (\ref{est1}) that $\dps \lim _{\e \to 0}
{\bf N}(\e ,\d ,n) =0$.  \eop

\subsection{Remarks}
\para{3.4.1.} It is an immediate consequence of continuous dependence on
initial conditions for solutions of
ODE's that the probabilistic negligibility of the set of ``bad'' initial values
${\cal B}:= \{ q_0 \in \g _0 : \tau
(q_0)<\infty \}$, $\P ({\cal B}) =0$,
implies the negligibility of $\cal B$ in the topological sense: $\cal B$ is
of first category in ${\g _0}$, i.e., it is contained in a countable union of
nowhere dense (in ${\g _0}$) sets.
(Take ${\cal B}_t = \{ q\in \g _0: \tau (q)\leq t\}$; cf.\ also \cite{Saari}.)
In other words: Global existence of Bohmian mechanics is typical and generic.
\para{3.4.2.} Since $\P$ is equivalent to the Lebesgue measure ${\bf L\/}$ on
$\g_0$, we have also that ${\bf L}({\cal B})=0$ and we thus have the global
existence and uniqueness of Bohmian mechanics ${\bf L}$-a.s. on $\g_0$.
\para{3.4.3.} The flux argument shows that any given hypersurface in
$\Omega\times\R$ (where $\psi$ is
$C^\infty$) of codimension greater than 1 will (almost surely) not be reached.
\para{3.4.4.}
We have shown that under certain conditions on the initial \wf\ and the
Hamiltonian, particle trajectories exist as solutions of (\ref{bohmev})
globally in time for $\P$-almost all initial conditions. In the
introduction we have already given an example showing that in general
(i.e., assuming merely the conditions of Theorem \ref{mainth} or Corollary
\ref{maincor}) this result does not hold for {\em all}\/ initial
configurations.  However, in that example the dynamics is uniquely
extendible to a global dynamics $Q:\R ^2 \to \R ,(q,t) \mapsto Q_t(q)$.
There are 3 continuous trajectories which periodically run into nodes of
the \wf, while the other trajectories are global solutions of
(\ref{bohmev}). This extended dynamics $Q_t(q)$ is continuous.

However, if the trajectory running through the node at $t=0,\,q=1$ is
analyzed, one finds that locally $Q_t(1)\sim\sqrt[3]{\frac34 t^2}+1$, i.e.,
the map $Q_t(q)$ is not differentiable with respect to $t$ at $t=0$ for
fixed $q=1$. This may, for example, be seen by considering the flux through
$q=1$ for $t$ near 0, or, what amounts to the same thing, by employing the
Formula (\ref{smap}) (see Section \ref{bmsa}) expressing the trajectories
as curves of constant value of the function $\dps F(q,t)= \int _{-\infty}^q
|\psi_t |^2\, dx$. (This behavior of trajectories hitting nodes is in fact
typical---though it does not occur in the example for the trajectory at the
origin; in fact, if $\psi(q,t^*)$ has a node of order $k$ at $q^*$,
$\psi(q,t^*)\sim\alpha x^k$ with $x=q-q^*,$ then $F(q,t^*) \sim
F(q^*,t^*)+ax^{2k+1},\, a=\frac{{|\alpha|}^2}{2k+1},$ and $\dps \parfrac
Ft(q,t^*) = -j^{\psi _{t^*}}(q)\sim bx^{2k},$ so that $F(q,t)\sim
F(q^*,t^*)+ax^{2k+1}+bx^{2k}s+cs^2,\, s=t-t^*,$ in the vicinity of the node.
Thus for $c\neq0$, the equation $F(q,t)=F(q^*,t^*)$ implies that
$x\sim\sqrt[2k+1]{-\frac cas^2}$.)

Concerning the regularity of  $Q_t(q)$ in $q$ at fixed $t$, one sees in the
example that for
suitable choices of initial time the solution map will fail to be
differentiable at $q=0$ (where there will be a fifth root singularity) or
at $q=\pm1$ (where there will be a cube root singularity) as a function of
$q$ for fixed $t$.

For an even stronger breakdown of regularity in $q$ for fixed $t$, consider
the harmonic oscillator in 3 dimensions, and take the $(n=1, l=1)$-state
$\psi (q,t) = r e^{-(r^2+z^2)/2} e^{i\phi}e^{-5it/2}$ in cylindrical
coordinates. This \wf\ vanishes only at $r =0$, i.e., on the $z$-axis.
Particles circle around the $z$-axis with angular velocity $1/r^2$. The map
$Q$ is uniquely extendable to a global dynamics given by a continuous map,
which is however not
differentiable with respect to $q$, by defining $Q_t(q_0)=q_0$ for all $t$
and $q_0\in \No _0$.

It is possible also to give an example in which the extended map must fail
even to be continuous with respect to $q$ for fixed $t$: Consider free
motion in 1 dimension, and let the \wf\ $\psi$ be even, (real and
positive), $C^\infty$, and supported on $[-b,-a]\cup [a,b]$ with
$0<a<b<\infty$. Then $\psi\in C^\infty (H_0)$. Moreover, there is a $t_1>0$
such that $\left(e^{i H_0t_1/\hbar}\psi\right)(0)\neq0$. Let
$\psi_0=e^{iH_0t_1/\hbar}\psi$.  $\psi _t$ is then even for all $t$,
so that the velocity field is odd, i.e., symmetric under reflection.  Any
extension $Q$ which respects this symmetry must have $Q_t(0)=0$ for all
$t$. Then the map $Q$ is discontinuous in $q$ for $t=t_1$, and, in fact,
any extension must have this discontinuity.

\para{3.4.5.} It is well known---at least if $V$ is real analytic in
$\Omega$ (see for example \cite{Rauch}, page 98)---that if $\psi$ vanishes
on a nonempty (bounded) open set in configuration-space-time, it vanishes
everywhere (in the components of $\Omega\times\R$ that intersect this set).
We remark that under the hypotheses of Corollary \ref{maincor}, the same
conclusion would in fact obtain merely if $\psi$ were to vanish everywhere
on the boundary of such a set (and even with the possible exception of a
single piece of the boundary contained in a constant-time hyperplane),
since it would then follow from global existence and the inaccessibility of
the nodes that $\psi$ must  vanish everywhere in this set.

\para{3.4.6.}
The probability of reaching the nodes $\P ( x\in (\partial {\cal
N}^\epsilon \cap \gdnt ))$ may also be estimated without using flux
integrals. We include this argument, which involves a choice for ${\cal
N}^\e$ different from the one used earlier. We remark that for the new
${\cal N}^\e$ we can see no reason why $\partial {\cal N}^\e$  must be
smooth, even piece-wise.  Notice also that Lemma \ref{nnodesl} involves
both stronger premises and, since the convergence in it is uniform, a stronger
conclusion than the corresponding Lemma \ref{sdnodesl}.
\begin{lemma} \label{nnodesl}
Assume {\rm A1--A4} and,  for $\epsilon >0$, let
\be{ndef} {\cal N}^{\epsilon} := \{(q,t)\in \Omega \times \R :\,
|\psi_t(q)|
\leq \epsilon\}. \ee
Then, uniformly in $\d$ and $n$, with  $x:= (Q_{\min (\tedl ,
T)},\min
(\tedl , T))$, \[  \lim _{\epsilon \to 0} {\bf P} (\{ q_0 \in \gedl _0: x\in
(\partial
{\cal N}^\e \cap \gdnt )\} ) =0 .\]
\end{lemma}
\bigskip

The proof involves a fairly standard ``existence of dynamics''
argument and is analogous to that of Nelson \cite{Nelson1} for the
similar problem in stochastic mechanics:
One looks for an ``energy'' function on the state space of the
motion which
becomes infinite on the catastrophic event. With good
a priori bounds  on the expectation value of that
function, one can control  the probability of catastrophic  events.

\proof  The function which recommends
itself here is $\log |\psi|$, i.e., what we control is the ``entropy.''

We first present a formal estimate, disregarding the problem that the
solution curve $Q_t(q)$ starting at $q$ may not exist
for all times---which is taken care of below.
Let $\bf E$ denote the
expectation with respect to ${\bf P}$.
We compute for arbitrary $T$:
\begin{eqnarray}
& & {\bf E}\left(|\log|\psi_T(Q_T)| - \log|\psi_0|\,|\right) =
{\bf E} \left ( \left | \int_{0}^{T}\frac{d}{d t}
\log|\psi_t(Q_t)|dt \right|
\right) \nonumber\\
& = & {\bf E} \left( \left|
{ \int_{0}^{T} \left(
\frac{1}{2}\frac{1}{|\psi_t(Q_t)|^{2}}\frac{\partial
|\psi _t (Q_t)|^{2}} {\partial t}  +  \frac{\left(\nabla |\psi
_t|\right)(Q_t)}{|\psi _t(Q_t)|}
\cdot v^{\psi_t}(Q_t) \right) dt } \right| \right)  \leq
\nonumber\\
& & \int_{0}^{T} {\bf E} \left(\frac{1}{2}\frac{1}{|\psi
_t(Q_t)|^{2}}
\left|\frac{\partial |\psi _t (Q_t) |^{2}}{\partial t}
\right|\right) dt  +
\int_{0}^{T} {\bf E} \left( \mu \frac{|\nabla \psi _t(Q_t)|^{2}}
{|\psi _t(Q_t)|^{2}}\right) dt , \label{epsi}
\end{eqnarray}
where we used for the inequality the bounds
\begin{equation} \nabla |\psi | \leq |\nabla \psi | \quad \mbox{
and } \quad
|\vpsi| \leq \mu \left|\frac{\nabla \psi}{\psi}\right|
\label{vpsibound}
\end{equation}
Now use the equivariance of $|\psi |^2$ (cf.\ (\ref{equivar}))
to compute the expectation
 ${\bf E} (f_t(Q_t))=\int _{\Omega} |\psi_t(q)|^{2}
(f_t(q))dq$ and obtain that the right hand side of
(\ref{epsi}) is equal to
\begin{equation}
\int_{0}^{T} \int_{\Omega} \frac 12 \left| \frac{\partial
|\psi_t(q)|^{2}}
{\partial t}
\right|dq \; dt + \mu \int_{0}^{T} \int_{\Omega} |\nabla
\psi_t(q)|^{2}
dq \; dt.  \label{15}
\end{equation}
By virtue of (\ref{fluxH}) we replace $| \partial |\psi_t(q)|^{2}
/
\partial t |$ by $|\psi^*_t(q) H \psi_t(q) - \psi_t(q) H \psi
^*_t(q) |/\hbar
$.
By Schwarz's inequality, the first term of (\ref{15}) is then bounded
by
\[ \frac 1\hbar \int_0^T \|\psi_t\|\,\|H \psi_t\| \; dt \ = \frac
T\hbar
\,\|H \psi_0\| <\infty , \]
and the second term is bounded for each $T<\infty $ by Assumption A4.

\bigskip
To construct from this  a rigorous proof we need only define a suitable
killed  process.
For $t\geq 0$ we define $\Qedlt : \gedl _0
\cup \{ \dag \} \longrightarrow
\gedl _t \cup \{ \dag \} $ by
\begin{equation}
\Qedlt (q):= \left\{\begin{array}{cl}
Q_t(q)     & \mbox {for } t\leq \tedl (q) \\
\dag       & \mbox {for } t > \tedl (q)
\end{array}
\right.           \label{flueps}
\end{equation}
For completeness, we set $\Qedlt
(\dag ) = \dag$ for all $t\geq 0$.
Consider the probability measure $P^{\epsilon \delta n}_0$ on
$\gedl _0 \cup \{ \dag \}$ which has the density
\[ \rho _0^{\epsilon \delta n} (q):= |\psi _0(q)|^2 \mbox { for }
q\in
\gedl _{0}\]
(and, of course, $ P _0^{\epsilon \delta n} (\dag ) = 1-\int
_{\gedl _{0}}
\rho _0 ^{\epsilon \delta n} (q) \; dq ). $
The image measure of the process \Qedlt \ is denoted by
$P^{\epsilon
\delta n}_t
:= P^{\epsilon \delta n} _0\circ (\Qedlt )^{-1}$ and has the
density
$\rho _t^{\epsilon \delta n}$ on $\gedl _t$. {}From the definition of
${\cal N}^\e$ (\ref{ndef}),
\be{star} \{ q_0 \in \gedl _0: x\in (\partial {\cal N}^\e \cap \gdnt )\}
\subset  \{ q_0 \in \gedl _0: |\psi (x) | = \e
\} .\ee
Since we keep $\delta$ and $n$ fixed, and since the estimates
are independent of $\delta$ and $n$, we will omit the indices
$\delta$ and $n$ on $\Qedl ,\gedl ,\roedl $.

Define for $q \in \ge _0 $ and $t \geq 0$
\[ D^{\epsilon}_t(q):=\log |\psi _{\min  (\te (q), t)} (
Q _{\min  (\te (q), t)} (q))| -
\log |{\psi}_0(q)|.\]  One has that
\[ D^{\epsilon}_T(q) =
\int_{0}^{T}\frac \partial {\partial  t} D^{\epsilon}_t(q) \,
dt = \int _0^T f_t \circ \Qetq \, dt, \]
where \[ f_t(y) := \left\{\begin{array}{cl}
0     & \mbox {for } y = \dag \\
\dps \frac 12  \frac{1}{|\psi_t (y) |^{2}}\frac{\partial
|\psi_t (y)|^{2}} {\partial t}
+ \frac{{\nabla}|\psi_t (y )|}{|\psi_t (y )|}
\cdot v^{\psi _t}(y )     & \mbox {for } y \in \ge _t
\end{array}
\right. \]
We shall show that uniformly in $\epsilon$
\begin{equation}
{\bf P}( \{q\in \ge _0 :
|D^{\epsilon}_T(q)| > K\} )\rightarrow 0
\; \mbox{as} \; K \rightarrow \infty . \label{pdeltae}
\end{equation}
Then, since for $q_0$ as in (\ref{star})
$D^{\epsilon}_T(q_0)=\log\e-\log|{\psi}_0(q_0)|$, the lemma follows from
(\ref{pdeltae}) by observing that
\[ {\bf P}(\{ q\in \ge _0 :\; | \log |{\psi}_0(q)|| > K\} )
\rightarrow 0 \; \mbox{as $K \rightarrow \infty$} \]
holds uniformly in $\epsilon$, which is immediate since the
density of ${\bf P}$ is $|\psi _0|^2$.

By Markov's inequality we obtain that
\begin{equation}
{\bf P}(\{q\in \ge _0 :
|D^{\epsilon} _T(q)| > K\} ) \leq
 \frac{1}{K} {\bf E} \left ( \charfct _{\ge _0}  \left|\int_{0}^T
f_t \circ \Qet \, dt \right| \right) . \label{markov}
\end{equation}
Recall now that ${\bf P} = P^\epsilon _0$ on $\ge _0$, and that
 $f_t =0 $ at $\dag$. Then by the definition
of $\rho _t ^\epsilon$ as the density of the image measure of
$\Qet$
 one obtains that the right hand side of (\ref{markov}) is bounded
by
\[  \frac{1}{K}  \int_{0}^T \int _{\ge _t} \roet (q)
| f_t (q)| dq \, dt \]
Using the bounds (\ref{rhobound}) (with $\rho _t$ replaced by $\rho _t
^\epsilon$) and  (\ref{vpsibound})
(which holds on $\g ^\e _t$) we finally obtain that
\begin{eqnarray}
\lefteqn{{\bf P}( \{q\in \ge _0 :
|D^{\epsilon}_T(q)| > K\} ) \leq}\nonumber\\
& & \frac{1}{K} \left(
\int_{0}^{T} \int _{\Omega} \frac{1}{2} \left| \frac{\partial |\psi
_t (q)|^{2}}
{\partial t}\right| dq \; dt +
\mu \int_{0}^{T} \int _{\Omega} \,
|{\nabla \psi _t(q)}|^{2} dq \; dt \right).        \label{36}
\end{eqnarray}
The bracket on the r.h.s.\ is (\ref{15}).
By the Assumptions A3 and A4, (\ref{15}) is finite and
hence
the r.h.s.\ of (\ref{36}) goes to zero uniformly in
$\epsilon$ as $K \rightarrow \infty $. Thus we have established
Lemma \ref{nnodesl}. \eop

\section{\BM \ and self-adjointness}\label{bmsa}
\para{4.1.} In this subsection we shall discuss the necessity of certain
assumptions under which we have established global existence of
the Bohmian particle motion (cf.\ Theorem \ref{mainth} and
Corollary \ref{maincor}). We shall investigate in particular the
assumptions concerning \SA\ of the Hamiltonian.

By Corollary \ref{maincor} we obtain global existence if the Hamiltonian is
the {\em form sum}\/ $H_0 + V$, and if the potential $V$ satisfies certain
conditions leading in particular to the Hamiltonian's being bounded from
below.  These conditions on the Hamiltonian guarantee in particular that
Assumption A4 of Theorem \ref{mainth} is satisfied. In the case of one
particle moving on the half line $\Omega = (0,\infty )$, we shall prove,
without invoking A4, global existence for a certain class of potentials for
arbitrary \sa\ extensions, which furthermore may be unbounded below.

\begin{theorem}\label{d1theor}
Let $\Omega = (0,\infty )$, ${\cal H}=L^2 (\Omega )$, and suppose $V\in
C^\infty
(\Omega )$ is such that $H_0+V$ is in the
limit point case at infinity (see for example \cite{Weidmann}). Let $H$ be an
arbitrary \sa\
extension of
$(H_0+V)|_{C_0^\infty (\Omega )}$, and let
$\psi_0\in C^\infty (H)$ with $\| \psi _0\| =1$. Then $\P (\tau
<\infty )=0$.
\end{theorem}

It follows for example from Theorem X.8 in \cite{ReeSiII} that if $V(r)\geq
-kr^2$ for $r>c$ with $c,k \geq 0$, then $H_0+V$ is in the
limit point case at infinity.

Consider as an example the potential $V(q)=-c/q^2$ with $c>0$ large enough:
The Hamiltonian $H= H_0+V$ is in the limit circle case at 0, in the limit
point case at infinity, and unbounded above and below (cf.\ for example
\cite{ReeSiII}).
Thus by Weyl's limit point-limit circle criterion there is a one-parameter
family of (similarly unbounded) \sa\
extensions of
$H|_{C_0^\infty (\Omega )}$ for all of which, by Theo\-rem \ref{d1theor},
\BM\ exists uniquely and globally for
$\P$-almost all initial values.

\bigskip

The proof employs a new definition of the particle dynamics in one
dimension which extends the solution to (\ref{bohmev}) and is interesting
in its own right.   (In fact, this definition extends the Bohm motion,
defined by (\ref{velfield}) and (\ref{bohmev}), to an equivariant motion
for all $\psi\in L^2$!)  Let $Q_t(q_0)$ be defined implicitly by
\[ \int_{-\infty}^{Q_{t}(q_0)} |\psi_t(q)|^{2} dq =\int_{-
\infty}^{q_0}
|\psi_0(q)|^{2} dq. \]
$Q_t(q_0)$ is well-defined if $$F(q,t) := \int _{-\infty}^{q}
|\psi _t|^2 dx$$ is strictly monotonic in $q$. This is
the case except  at extended intervals with $\psi_t =0$, where
$F(\cdot,t)$ has a plateau. To define $Q_t(q_0)$
globally for $q_0\in \R$, set for example
\be{smap} Q_t(q_0):= \min \{ q:F(q,t)=F(q_0,0)\} \ee
(and $Q_t(q_0)=-\infty$ if $F(q_0,0)=0$, $Q_t(q_0)=\infty$ if $F(q_0,0)=1$).

\proof {}From Lemma \ref{regl} we obtain that $\psi \in C^\infty (\Omega
\times \R )$. Therefore, using the continuity of the scalar product and the
$L^2$-differentiability
of $t\mapsto \psi_t$,
\[ F(q,t)=\int _0^q |\psi _t|^2 \, dx = (\charfct _{[0,q]}\psi_t
,\psi_t ) \]
(where $(\cdot,\cdot)$ denotes the scalar product in ${\cal H}=L^2 (\Omega
)$)   is jointly continuous and differentiable. Clearly $F(0,t)=0$, $\lim
_{q\to \infty }F(q,t)=1$, and $\partial F/\partial q = |\psi_t (q)
|^2$. Moreover,
\[ \frac{\partial F(q,t)}{\partial t} = \int_0^q \frac{\partial
|\psi _t|^2}{\partial t} \, dx = -j_t(q) + \lim _{c\to 0}j_t(c) = -j_t(q).
\]
Here the existence of $\lim _{c\to 0}j_t(c)$ follows for $\psi \in
C^\infty (H)$ from partial integration of $\int _c^d (\psi ^* (H\psi )-
(H\psi ^*) \psi) \, dx$ and Schwarz's inequality; the value 0 for
$\lim _{c\to 0}j_t(c) =0$ follows from the symmetry of $H$
together with the fact that $\lim _{d\to \infty}j_t(d) =0$, which holds because
$H$ is in the
limit point case at infinity. (See for example \cite{Weidmann}.)

For all $t$ and all $q_0 \in \g_0 = \Omega \setminus \No _0$,
let $Q_t(q_0)$ be defined by (\ref{smap}). It follows from the implicit
function theorem that $t\mapsto
Q_t(q_0)$ is continuous and differentiable for
$(q_0,t)$ such that $\psi_t (Q_t(q_0))  \neq 0$, with $dQ_t/dt=
j(Q_t)/|\psi_t (Q_t)|^2= v^{\psi _t}(Q_t)$, i.e.,
$Q_t$ solves the
differential equation (\ref{bohmev}) on ${\cal G}= (\Omega\times \R)
\setminus \No$.
It remains to show that for $\P$-almost all initial $q_0$, $\tau (q_0) =
\sup \{ s>0: Q_t(q_0) \in {\cal G}
\mbox{ for all } t\leq s\} $ is infinite, i.e., (\ref{bohmev}) has
global solutions for almost all initial values.

Now it is obvious from this definition that $Q_t(q_0)\in \Omega$ for all
$t$ and all $q_0\in \g _0$. ($Q_t (q_0)=0$ corresponds to $F(q_0,0)=0$,
$Q_t (q_0)=\infty$ to $F(q_0,0)=1$, and for $q_0 \in \g_0$, $F(q_0,0)\in
(0,1)$.) Moreover, by the $L^2$-continuity of $t\to\psi_t$, we have that
for $0<T<\infty$ and $q_0\in\g_0,\ \inf_{0\leq t\leq T} Q_t (q_0)>0$ and
$\sup_{0\leq t\leq T} Q_t (q_0)<\infty$, i.e., the trajectories cannot run
into the (only) possible singularity of the potential ${\cal S}=\{ 0\} $ or
to infinity in finite time. Thus it remains only to control the probability
of hitting ${\cal N}$, for which Lemma \ref{sdnodesl} does the job. We
omit the details.   \eop

\para{4.2.} One might now wonder whether we have global existence of \BM\
for any \sa\ Schr\"odinger Hamiltonian (without assuming A4). This is quite
trivially wrong, as is easily seen by considering  free motion on the
interval $\Omega = (0,1)$. There are
\sa\ extensions of $H_0|_{C_0^\infty (\Omega )}$ with
$j(0)=j(1)\neq 0$.  (Similarly one might consider potentials on $\Omega =
(0,\infty )$ such that $H_0 + V$ is in the limit circle case at infinity.)
This corresponds to an incoming flow at 0, balanced by an outgoing flow at
1 (or the other way round) so that the total probability is conserved (a
situation which can of course be identified with a motion on a
circle). Typically, the particle will reach the boundary of $\Omega$, so that
almost sure global existence in the sense of solutions of the differential
equation (\ref{bohmev}) fails.  However, the motion is quite trivially
extendible in such a way that the trajectories are piecewise solutions of
the differential equation: when the boundary of $\Omega$ is  reached they
jump to the other end of $\Omega$. $|\psi |^2$ then remains an
equivariant measure. This motion can be described by replacing (\ref{smap})
by
\[  Q_t(q_0):= \min \{ q:\wt F(q,t)=\wt F(q_0,0)\} \] with
\[ \wt F(q,t) = \left( F(q,t) - \int _0^t j_s(0) \, ds \right)  (\mbox{mod }1)
\]
[Another possibility to define a global motion in this case is to use the
unmodified (\ref{smap}). This provides then an example of a deterministic
dynamics completely different from (and not an extension of) the Bohmian
dynamics, ((\ref{bohmev}) is replaced by the nonlocal form $dQ/dt = (j^\psi
-j_t(0))/|\psi |^2$) for which, however, $|\psi |^2$ remains  equivariant.
With this motion, particles do not jump from 1 to 0 or the other way
round. (However, they might all run through nodes!)]

\bigskip

In fact, we expect generally    that \SA\ guarantees
(possibly discontinuous) extendibility
of the Bohmian motion in such a way that $|\psi |^2$ is an
equivariant
measure. This is suggested by the fact that the symmetry of the
\ham\
leads to
\[ \lim_{\delta\rightarrow 0, n\rightarrow \infty} \left(
 \int_{\partial {\cal S}^{\delta} \cap {\cal K}^n }
(j^{\psi_t}(q)\cdot u) ds \, \, + \,\,
\int_{\partial {\cal K}^{n} \setminus {\cal S}^\delta }
(j^{\psi_t}(q)\cdot u) ds \right) =0, \]
using integration by parts (Green's identity)
\[ \int_{M} \psi^{\ast} (H\psi) dq - \int_{M} (H \psi^{\ast}) \psi
dq
\;=\;-i\hbar \int_{\partial M} (j^{\psi}\cdot u) ds , \]
for $M= {\cal K}^n \backslash {\cal S}^{\delta}$.
The vanishing of the integrals over the {\em absolute}\/ flux
yields global existence of \BM : In
finite time the singularities and infinity are not reached. The
flux balance from \SA\ alone suggests extendibility of the motion:
Some parts of the singularities (or infinity) may act as sources, others as
sinks.

\para{4.3.} For a wider perspective on this matter let us consider a \Schr\
Hamiltonian $H$ on a domain
where it is not (essentially) \sa , i.e., where the boundary conditions are
too few or too weak. Then, first of all, the time evolution of \wfs\ is not
unique: There are infinitely many different unitary evolutions
(corresponding to the different \sa\ extensions), and there are also
semi-groups for which $\| \psi _t \|$ is not conserved.  The (essential)
\SA\ of $H$ is equivalent to ${\rm Ker} (H^*\pm i)= \{ 0\}$, so that if $H$
is considered on a domain where it is symmetric but not \sa , then $H^*$
has imaginary eigenvalues. Together with the (space) regularity for
eigenstates of the elliptic operator $H^*$ (assuming sufficient regularity
for the potential $V$) we thus obtain classical solutions of \Schr 's
equation with exponentially decreasing or increasing norm. Since $\rho =
|\psi |^2$ still holds on $\cal I $ (cf.\ the paragraph around Equation
(\ref{rhobound})), those solutions lead with positive probability to
catastrophic events.

This possibility is not that far-fetched: The \ham\ for one particle in a
Coulomb field $V(r)=-1/r$ considered on the ``natural'' domain $C_0^\infty (\R
^3\setminus \{ 0\} )$ is not essentially self-adjoint and hence the time
evolution of the \wf\ is not uniquely defined \cite{Kemble,JR}. There are
many properties that mathematically distinguish the \sa\ extension usually
regarded as ``the Coulomb \ham '' from other possible extensions.  However,
we do not know of any convincing (a priori) physical argument for ``the
Coulomb \ham '' unless one accepts, for example, that the Coulomb potential
is a ``small perturbation'' of the free Hamiltonian \cite{kato51}, or that
``in reality the singularity is smeared out.''  Of course, if we require that
\BM\  be globally existing, then, as we have argued above, only \sa\
extensions are possible. But among all
\sa\ extensions \BM \ seems not to discriminate: While our Corollary
\ref{maincor} applies only to the form sum
(which is ``the Coulomb \ham ''), it is heuristically rather clear (or at
least plausible) that
\BM\ should exist globally and uniquely for all the other \sa\ extensions of
$H|_{C_0^\infty (\R
^3\setminus \{ 0\} )}$ as well.\footnote{The singularity of the radial
current at 0 may be estimated for $\psi = \sum _{l,m} f_{lm}(r)
Y_{lm}(\theta ,\phi )\in \bigoplus _{l=0}^\infty {\cal D}(H_{r,l})\otimes
K_l$, where, for angular momentum $l$, $H_{r,l}$ is the radial part of
$H$ and $K_l$ is the corresponding eigenspace of the angular part of
$-\frac{\hbar ^2}{2m}\Delta$, as
follows: $H_{r,l}f_{lm} \in L^2(\R ^+, r^2\, dr)$ implies for the worst
behavior of $f_{lm}$ as $r\to0$ that $f_{lm}\sim r^\alpha$ with $\alpha
>1/2$ for $l\geq 1$ resp.\ $\alpha =-1$ for $l=0$. Therefore the worst
behavior of the the radial current $j_r^\psi = \frac{\hbar}m{\rm Im} \left(
\psi ^*
\frac{\p \psi}{\p r}\right)$ as $r\to 0$ is $j_r^\psi \sim r^{-(3/2) +\e } $
(using the fact that the radial current at 0 vanishes on ${\cal D}(H_{r,0})$
for all \sa\ extensions, cf.\ the
proof of Theorem \ref{d1theor}). Thus we should have that $\int _{{\cal
K}^r}|j^\psi \cdot u|
\, ds = \int _{S^2}|j^\psi _r |\, r^2\, d\omega \sim r^{(1/2) +\e } \to 0$ as
$r\to 0$. For a proof of global existence along the lines of Theorem
\ref{mainth}, it is necessary also to control the time change of the radial
current. However, the global existence for the one-dimensional problem
(Theorem
\ref{d1theor})  suggests that this should be possible.}

Nonetheless, discussions about the ``right''
(unitary or contractive) evolution, i.e., about the ``right'' boundary
conditions, as for example in the case of
strongly singular potentials
like the $1/r^2$ potential (see \cite{Case,Landau,Nelson2}), do
gain now firm ground by taking into account the
actual behavior of the particles:
Whether or not we should consider the Bohmian particle to be
caught at the origin is a matter of the physics we wish to describe:
whether or not particles disappear in the nucleus.
An axiom, or dogma, of self-adjointness of the Hamiltonian (or
equi\-va\-lent\-ly of unitarity of the \wf\ evolution) appears quite
inappropriate
from a Bohmian perspective---even though the importance of self-adjointness
is profoundly illuminated by this perspective!

Moreover, the particle picture of \BM\ naturally yields an interpretation
of the current $j$ as a current of particles moving in accordance with the
density $|\psi |^2$. In this way, boundary conditions for \SA\ of the form
$j=0$ at the singularities or $j(in)=j(out)$ may be viewed as ``arising
from \BM .'' For example, the outcome of a detailed analysis of \sa\
extensions of $H_0$ on the half line $(0,\infty )$---there is a one
pa\-ra\-me\-ter family of \sa\ extensions $H_0^a$, the respective domains being
defined by $\psi '(0)/\psi (0) =a$, $a$ real, or $\psi (0)=0$
($a=\infty$)---is easily guessed from the point of view of \BM \ by
demanding that either $\vpsi (0) =0$, i.e., Im$(\psi '(0)/\psi (0)) =0$, or
that $|\psi (0)|^2=0$.

\para{4.4.} We wish to conclude with some remarks on the general Hilbert
space description of orthodox quantum theory viewed from  the perspective
of \BM.
We have discussed the fact that \BM\ is well defined, i.e., trajectories
exist
globally and uniquely, for typical initial values and for \wfs\
which
are $C^\infty$-vectors of the
self-adjoint Hamiltonian $H$. The set $C^\infty (H)$ is dense and
invariant;
however, it is most likely not a residual in the norm topology of
the Hilbert space
$L^2(\Omega )$, i.e., it is presumably not a ``generic'' set, and
it furthermore depends on the Hamiltonian.
One might now wonder how \BM\ can be taken as the
basis for the quantum formalism (as has been claimed---see
\cite{DDGZ})
if the former cannot even be defined
for a really ``fat'' set of \wfs. And since, as we have seen, \BM\
yields  a natural understanding of the (spirit of the) meaning of the \SA\
of a Schr\"odinger Hamiltonian, the question
should be even more puzzling. The answer is, of course, that the embedding of
\BM\ into a Hilbert space
structure is a natural but purely mathematical device.
Indeed this answer is (of course, in disguise) commonly
accepted---though maybe not as loudly stated: No physicist
believes
that a generic $L^2$-\wf\ (in the residual sense) results as
the ``collapsed''
\wf\ from a preparation procedure. The state space  of physical \wf s
$\psi$
is not the Hilbert space ${\cal H}=L^2(\Omega )$ but more or less the space of
classical,
smooth solutions of Schr\"odinger's equation, for the analysis of which the
$L^2$-norm and hence the Hilbert space structure is of critical importance.

Other aspects of this embedding are commonly taken more seriously: for
example, that observables are self-adjoint operators on $\cal H$. While we
do not wish here to enter into a general discussion of this question
(see \cite{DDGZ}), we would like once again to comment on the
self-adjointness of the Hamiltonian $H$. The importance of this property is
certainly not that ``measured energy values must be real'' but lies rather in
Stone's theorem: $H$ acts as the generator of a one-parameter unitary group
$U_t$, which gives the time evolution of states $\psi_t = U_t \psi_0$ (or
 of observables $A_t = U_t^{-1} A U_t$), and hence must be self-adjoint
by Stone's theorem.  Why should the time evolution be unitary?  Simply
because the norm $\|
\psi_t \|$ must be invariant, so that the total probability is
conserved.

We conclude with some remarks about effective descriptions.  We first note
that restrictions of configuration space such as described in the last
paragraph of Section 4.3 (with a freedom in the boundary condition) are
perhaps best understood physically as arising as a limit of a sequence of
(moderately realistic) potentials $V^a_n$ tending to ``$V=0$ for $q>0$,
$V=\infty$ for $q<0$'' in a suitable way---such that $H_0+ V^a_n \to H^a_0$
in an appropriate sense.  This problem is analyzed in \cite{Seba,Alb,Lara},
and in \cite{Lara}  the convergence of the Bohmian trajectories in this limit
is derived.

In other physically interesting but complex situations we may have an
effective description involving a \ham\ which is self-adjoint but not of
Schr\"odinger-type,\footnote{The modeling of physical situations leads
often to idealizations which are very singular. In Newtonian mechanics one
considers for example singular evolutions induced by ``hard walls''
confining a particle or by elastic collision between hard spheres.} so that
the probability current $j$ may fail to be of the usual form [cf.\ Eq.\
(\ref{qflux})], or where there may in fact be no local conservation law at
all for the probability density ${|\psi|}^2$.

For an example with nonstandard current $j$, consider the self-adjoint
shift operator $H_{c}=-i\hbar c \nabla $, where $c$ is a constant with the
dimension of a velocity.  $\psi _t(q) = e^{-itH_{c}/\hbar }\psi _0(q)=\psi
_0 (q-ct)$ describes translation without ``spreading.'' This Hamiltonian
may perhaps arise in a limit in which the spreading
of the \wf , induced by the Laplacian, can be neglected. In any case, the
corresponding current is $j_c=c{|\psi|}^2$, and the obvious candidate for
the ``Bohm motion'' in this case is $v=j_c/\rho=c$, not (\ref{velfield}).

It is conceivable
that an approximation procedure leading to an effective Hamiltonian like
$H_c$, when applied to
\BM , also converges to a deterministic limit. If so, then $v^{\psi}=c$
would be the natural  guess for the motion in this limit.

For a \ham\ with no local conservation law for probability there is of
course no ``Bohm motion'' generalizing (\ref{velfield}).

\section{Acknowledgements}
During this work, we profited from discussions with many collegues. In
particular, we wish to acknowledge helpful discussions with Lara Beraha,
Martin Daumer, Joseph Gerver, John Mather,  Markus Schneider, Eugene Speer, Avy
Soffer, Herbert Spohn,
Francois Treves, and J\"urgen Weckler.
This work was supported  by the DFG, by
NSF Grant  No. DMS-9305930, and by the INFN.

\section{Appendix: On the regularity of $\psi$}
\begin{lemma} \label{regl}
Assume {\rm A1--A3}, and let $\psi _t(q)= e^{-itH/\hbar} \psi
_0(q)$. Then
there exists a function
$\wt \psi \in  C^\infty (\Omega \times \R )$ such that for all
$t\in \R$
$\wt \psi (q,t) = \psi _t(q) $ for almost all $q$. ($\wt \psi$ is a
classical solution of Schr\"odinger's equation.)
\end{lemma}

This fact is presumably folklore knowledge to experts
in PDE's, but since we could find no suitable reference---and since it does
not appear to be well known among mathematical physicists---we shall supply a
proof. (Hunziker \cite{Hunzi} has established space-time regularity of $\psi$
for
potentials which are bounded, have bounded derivatives, and are $C^\infty$ on
$\R ^d$ for $\psi _0 $ in
Schwartz space. Also, regularity (in space) of eigenfunctions (for sufficiently
regular potentials) is well known
\cite{ReeSiII}.)

\proof  We apply standard methods of elliptic
regularity (see, for example, \cite{rudin}) to the elliptic
operator $L$ on $\Omega \times \R$
\[ L := -\hbar ^2 \frac {\partial ^2}{\partial t^2} - \sum
_{k=1}^N
\frac {\hbar ^2}{2m_k} \Delta _k +V = -\hbar ^2 \frac {\partial
^2}
{\partial t^2} +H \]  {}From A3,
$\psi _t \in C^\infty (H)$, and therefore the functions $ \phi
_{n,t} :=
H^n \psi _t (=e^{-itH/\hbar }\phi _{n,0})$ are in $L^2(\Omega )$
for
all $n$ and $t$. With this definition, formally
\begin{equation} \label{L}
L\psi = \phi _2 + \phi _1, \quad \mbox{and} \quad L\phi _n =
\phi _{n+2} +  \phi _{n+1} \end{equation}
To apply the theorem of elliptic regularity, we need to show a)
that
$\psi$ and $\phi _n$ are locally $L^2$ in $\Omega \times \R$,
therefore
locally in the Sobolev space $W^0$ (we refer to the definitions
and
theorems of \cite{rudin}, where however $W^n$ is written $H^n$) in $\Omega
\times \R$, and b)
that (\ref{L}) is  satisfied in the distributional sense on $\Omega
\times \R$. Then, by repeated use of Theorem 8.12 in \cite{rudin}
we obtain
that $\psi$ (and $\phi _n$) are locally in $W^n$
for all even (positive) integers $n$. Then by Sobolev's lemma
$\psi$ is
indeed (almost everywhere equal to) a $C^\infty$-function on
$\Omega \times \R$. (The space-time set of measure 0 on which
$\psi$ has
to be corrected indeed splits into $t$-slices that are of measure
0 for
all $t$. This is a consequence of $L^2$-continuity of $t\mapsto
\psi _t$.)

a)  This is an easy consequence of Fubini's theorem if
$\psi$ and $\phi _n$ are jointly measurable in $q,t$. $\psi_t$ is
measurable in $q$ ($\psi_t \in L^2(\Omega )$) and
the map $t \mapsto \psi _t$ resp.\  $t \mapsto \phi _{n,t}$
is weakly
measurable (indeed much more is true, namely strong
differentiability).
Then by a theorem of Bochner and von Neumann \cite{bvn} joint
measurability
of $\psi_t(q)$ and $\phi _{n,t}(q)$ in $(q,t)$ follows in the
following sense:
There exist functions $\wt \psi$ and $\wt \phi _n$ which are
jointly measurable in $q,t$, and for all $t$
$\wt \psi (q,t)=\psi _t(q)$ and $\wt \phi _n(q,t)=\phi _{n,t}(q)$
for almost all $q\in \Omega $. In the following, we shall denote
$\wt \psi (q,t)$ and $\wt \phi _n (q,t)$  by $\psi (q,t)$ and
$\phi _n (q,t)$ or again by $\psi _t(q)$ and $\phi _{n,t}$, as
convenient.

b) First one convinces oneself that $\psi$ and $\phi_n$ satisfy
Schr\"odinger's
equation in the distributional sense, i.e., for all test functions
$f\in C_0^\infty (\Omega \times \R)$,
\[ -i\hbar \int (\frac {\partial}{\partial t} f) \psi \ dq\ dt =
\int (Hf)\psi \ dq\ dt .\]
This follows by looking at the function $G:\R \to \R ,t\mapsto
\int f(q,t)  \psi (q,t) dq = (f_t^*,\psi _t)$, where
$(\cdot,\cdot)$ denotes the scalar product in $L^2(\Omega )$. $G$ has
compact
support, and its derivative is seen to be
\[ \frac {dG(t)}{dt} = \frac 1{i\hbar}
(f_{t}^*, H\psi _{t}) + (  \frac {\partial f_t^*}{\partial t} ,
\psi _{t})\]
by the continuity of the scalar product $(\cdot,\cdot)$ and the weak (in
$L^2(\Omega )$) differentiability of $\psi_t$. The same holds with $\psi$
replaced by $\phi _n$ for all $n$.  Since furthermore $f_t \in {\cal D}(H)$
for all $t$, and therefore $(f_t^*,H\psi _t) = (Hf_t ^*,\psi _t )$, $\psi$
and $\phi _n$ are indeed weak solutions of Schr\"odinger's equation (and
the self-adjoint operator $H$ on ${\cal D}(H)$ agrees with the operator on
distributions defined by $H$), i.e., we've arrived at $$i\hbar \frac {\partial
\psi}{\partial t} = H\psi =\phi _1\quad
\mbox{and} \quad  i\hbar \frac {\partial \phi _n}{\partial t} =
H\phi _n
=\phi _{n+1}$$ weakly, and therefore
(\ref{L}) indeed holds in the distributional sense (on $\Omega \times
\R$).\eop

\end{document}